\documentclass[11pt]{article}
\usepackage{amssymb}
\usepackage{amsmath,amsthm}
\usepackage{enumerate}
\usepackage{graphicx}
\usepackage{subfigure} % for multiple figures together
\usepackage{setspace}
\usepackage[left=25mm,top=24mm,right=25mm,bottom=18mm]{geometry}

\allowdisplaybreaks

\newtheorem{prop}{Proposition}
\newtheorem{theorem}{Theorem}
\newtheorem{lemma}{Lemma}
\newtheorem{Definition}{Definition}

\theoremstyle{remark}\theoremstyle{plain}
\theoremstyle{definition}%
{Assumption}\theoremstyle{plain}

\newenvironment{assumptiontwo}[2]{\par\noindent\ignorespaces\textbf{Assumption #1} (#2)}{\par\noindent\ignorespacesafterend}

\def\R{\mathop{\mathbb R}\nolimits}

\newcommand{\E}{\mathop{{}\mathbb{E}}\nolimits}

\newcommand{\LL}{\mathcal{L}} % to denote L matrix

\title{
\vspace{-2.0cm}% to keep everything in a single page
{\small
\begin{minipage}{.49\linewidth}
\begin{flushleft}
{https://doi.org/10.1016/j.jeconom.2021.07.002}
\end{flushleft}
\end{minipage}
\begin{minipage}{.49\linewidth}
\begin{flushright}
Journal of Econometrics, accepted manuscript.\footnote{
\copyright~2021 This manuscript version is made available under the CC-BY-NC-ND 4.0 license
{https://creativecommons.org/licenses/by-nc-nd/4.0/}
}
\end{flushright}
\end{minipage}
}
\\
\vspace{1.0cm}
Fully  Modified Least Squares Cointegrating Parameter Estimation in 
	 Multicointegrated Systems%\footnote{}
      }

\author{Igor L. Kheifets$^{a,b}$
\and Peter C. B. Phillips$^{c,d,e,f}$} 
\date{
$^{a}$HSE University, Russia \\  
$^{b}$Center for Econometrics and Business Analytics, St. Petersburg State
University, Russia\\
$^{c}$Yale University, USA\\
$^{d}$University of Auckland, New Zealand\\
$^{e}$University of Southampton, UK\\
$^{f}$Singapore Management University, Singapore\\[2ex]
	\today
} 
\begin{document}
\maketitle
\begin{abstract} \setstretch{0.95}
Multicointegration is traditionally defined as a particular long run
relationship among variables in a parametric vector autoregressive model
that introduces additional cointegrating links between these variables and partial sums of the
equilibrium errors. This paper departs from the parametric model, using a
semiparametric formulation that reveals the explicit role that singularity
of the long run conditional covariance matrix plays in determining
multicointegration. The semiparametric framework has the advantage that
short run dynamics do not need to be modeled and estimation by
standard techniques such as fully modified least squares (FM-OLS) on the
original $I\left( 1\right) $ system is straightforward. The paper derives FM-OLS limit theory in
the multicointegrated setting, showing how faster rates of convergence are
achieved in the direction of singularity and that the limit distribution
depends on the distribution of the conditional one-sided long run covariance estimator used in
FM-OLS estimation. Wald tests of restrictions on the regression coefficients have nonstandard limit theory which 
depends on nuisance parameters in general. The usual tests are shown to be conservative when the restrictions are isolated to the directions of singularity and, under certain conditions, are invariant to singularity otherwise. 
Simulations show that approximations derived in the paper work well in finite
samples. The findings are illustrated empirically in an analysis of fiscal sustainability of the US
 government over the post-war period. 

\vspace{0.08in}

\vspace{0.08in}

\end{abstract}

%\vspace{-8mm}

%\pagebreak

% ----------------------------------------------------------------
\renewcommand{\baselinestretch}{1.15}\normalsize
\section{Introduction}
Many economic time series are non-stationary and contain stochastic trends,
which are naturally modeled using cointegration. For example, two $I(1)$
variables $y_t$ and $x_t$ are \emph{cointegrated} if for some $A$, $u_{0t}=y_t-Ax_t$ is $I(0)$. 
Granger and Lee (1990) call \emph{multicointegration} a situation when the
cumulative error $U_{0t}=\sum_{s=1}^t u_{0s}$ is cointegrated with $x_t$ or
$y_t$. They analyze a case where $(y_t,x_t,u_{0t})$ are production, sales and
inventory investment, $A=1$ and $U_{0t}$ is the level of
inventories. Inventory stock $U_{0t}$ may then be cointegrated with production via an adjustment mechanism that captures firm decision making on inventory investment, as well as satisfying an identity arising from the aggregation of the defining relationship $y_t=x_t+u_{0t}$. 

It is important to take into account the presence of multicointegration in a cointegrated system: 
on one hand it can invalidate usual procedures of estimation
and testing in cointegrated systems by affecting asymptotic properties; and on the other it may lead to advantages in improved forecasting performance.
Multicointegration has so far been analyzed only in a VAR framework\footnote{
It is of course possible to write stationary VARs in moving average form and vice versa under invertibility conditions. There is now a large literature describing such explicit representations for cointegrated time series. For a general approach based of Laurent series representations, see Franchi and Paruolo (2019) and the references therein.}
and naturally involves implicit
restrictions on the model induced by the extra layer of cointegration. Engsted and Johansen (1999), for example, 
show that if the process is generated by a VAR model for $I(k)$ variables,
multicointegration may occur if $k=2$ but not if $k=1$. 
Likelihood-based estimators of cointegration parameters in $I(2)$ VAR
multicointegrated systems have mixed normal limit distributions and
likelihood ratio statistics for hypothesis testing about the parameters generally have asymptotic $\chi^2$ null distributions under conditions of correct specification,
as shown for example in  
Johansen (1997, 2006), Boswijk (2000, 2010), and Paruolo (2000). Berenguer-Rico and Carrion-i-Silvestre  (2011) provide an  application of this approach that examines government debt sustainability. 

In contrast to these studies, the present paper studies $I(1)$ cointegrated models that are possibly multicointegrated in a
semiparametric framework with specific focus on the use of fully modified least squares (FM-OLS) estimation. In related work, the authors (Phillips and Kheifets, 2019) explore the concept of multicointegration in a general $I(1)$ triangular cointegrated system
with weakly dependent errors, showing how multicointegration emerges naturally from singularity of the long run covariance matrix. This formulation gives an explicit mechanism generating multicointegration as a property of a general triangular $I(1)$ system, as opposed to imposing multicointegration subsequently on a parametric system like a VAR. The contrast lies in the capacity of a general I(1) triangular system to implicitly involve the effects of multicointegration without changing or restricting the cointegrating coefficients. The implicit effects propagate from the nonparametric treatment of the equation errors and are therefore typically unknown to the investigator. This property is one of the primary motivations of our study. A second motivation is to show that FM-OLS estimation of the cointegration coefficients has some useful robustness properties to the possible unknown presence of multicointegration. 

More specifically, the present paper contributes by developing asymptotic theory for FM-OLS estimation and testing in cointegrating relationships that involve multicointegration in a semiparametric setting. The analysis of triangular cointegrated systems under singularity that is developed is of some independent interest. The results show that cointegrated system estimation may proceed under certain conditions in a general $I(1)$ cointegrated system in the presence (and without prior knowledge) of multicointegration.  

To define multicointegration for weakly dependent data, we take the triangular
representation of a linear cointegrating relationship. In the cointegrating
regression model
\begin{align} \label{pp0}
y_t ={}& Ax_t+u_{0t},\;\; x_t =x_{t-1}+ u_{xt}, \quad t=1,\ldots,T, 
\end{align}
$A$ is a $m_0\times m_x$ cointegrating coefficient matrix, $x_t$ is initialized at $t=0$ by $x_0 =O_p(1)$,
and the combined error vector $u_t = (u'_{0t}, u'_{xt})'$ follows the linear process
\begin{align} \label{pp1}
u_t =D(L)\eta_t =\sum_{j=0}^{\infty} D_j\eta_{t-j},
\quad \eta_t\sim iid(0,I_m),
\quad\text{with }
\sum_{j=0}^{\infty} j^{\nu}||D_j|| <\infty,
\end{align}
for some $\nu>2$, finite fourth order cumulants of $\eta_t$, and
where $m = m_0 +m_x$. It is common in the literature to consider such time
series with an additional assumption $|D(1)|\ne 0$ (e.g. Phillips, 1995) that assures nonsingularity of the long run variance matrix of $u_t$, which
we relax here. 

Let $\Gamma_{u,u}(h)=\E u_{t+h} u'_{t}$. The linear operator $D(L)$, the long run covariance matrix
$\Omega=\sum_{h=-\infty}^{\infty}\Gamma_{u,u}(h)= D (1) D
(1)'=\sum_{k=0}^{\infty}\sum_{j=0}^{\infty}D_j D'_k$ of $u_t$ and  one-sided
long run covariance matrix $\Gamma^{+}=\sum_{h=0}^{\infty}\Gamma_{u,u}(h)=
\sum_{k=0}^{\infty}\sum_{j=0}^{k}D_j D'_k$ of $u_t$ are partitioned conformably
with $u_t$ as 
\begin{equation*}
D(L)=\left[
\begin{array}{cc}
D_{00}(L) & D_{0x}(L) \\
D_{x0}(L) & D_{xx}(L)%
\end{array}%
\right],
\Omega=\left[
\begin{array}{cc}
\Omega_{00} & \Omega_{0x} \\
\Omega_{x0} & \Omega_{xx}%
\end{array}%
\right],
\Gamma^{+}=\left[
\begin{array}{cc}
\Delta_{00} & \Delta_{0x} \\
\Delta_{x0} & \Delta_{xx}%
\end{array}%
\right],
\end{equation*}
where $\Omega_{xx} > 0$ is positive definite so that $x_t$ is a full rank $I(1)$ regressor vector, as commonly assumed in triangular systems such \eqref{pp0} and \eqref{pp1} following Phillips(1991)\footnote{The case where the regressors $x_t$ are themselves cointegrated is considered in Phillips (1995) but is not considered in this paper.}. The conditional long run
covariance matrix, defined as the Schur complement of the block $\Omega_{xx}$, is
$\Omega_{00.x}=\Omega_{00}-\Omega_{0x}\Omega^{-1}_{xx}\Omega_{x0}$ and is positive
(semi-) definite if and only if $\Omega$ is positive (semi-) definite (by virtue of the Guttman
rank additivity formula). 
In this paper we consider a situation when the long run variance
matrix is singular, or, equivalently, 
when the conditional long run covariance
matrix is singular.
It corresponds to a case where partial sums of $y_t$ and $x_t$ are cointegrated
with an $I(0)$ error in some unknown direction, i.e. when there is
a multicointegration in the spirit of Granger and Lee (1990), but is semiparametric in the sense that the short run dynamics are left unspecified. 
We therefore introduce the following definition. 
\begin{Definition}{}
The process generated by a triangular cointegrating system is called multicointegrated if its long run error covariance matrix is singular.
\end{Definition}

The advantage of this framework is that it provides the
explicit origin from which the multicointegrating relationship arises in an $I(1)$ system. Thus, if we take
partial sums of the augmented regression form (Phillips, 1991)
\begin{equation}\label{ppA}
y_t = A x_t +F (1-L) x_t + u_{0.x,t},
\end{equation}
where $F=\Omega_{0x}\Omega_{xx}^{-1}$ is the long run regression coefficient of
$u_{0t}$ on $x_t$ and $u_{0.x,t}=u_{0t}-\Omega_{0x}\Omega_{xx}^{-1}u_{xt}$,
giving (using capitals with time index $t$ for partial sums)
\begin{equation}\label{eq:xx}
Y_t = A X_t +F x_t + U_{0.x,t}.
\end{equation}
It becomes clear that in the direction of singularity of $\Omega_{00.x}$ we have an
exact long run relationship that links the time series $Y_t$, $X_t$, and $x_t$ and this relationship is prescribed in terms of the coefficients $A$, $F$ and the singular direction of $\Omega_{00.x}$, which
is estimable. In earlier work on multicointegration, the hypothesis about multicointegration
is imposed a priori, directly and explicitly as in Granger and Lee (1990) or through rank conditions in VAR analyses.
What our approach does is: 
(i) show that multicointegration may exist in a triangular $I(1)$ system such as \eqref{pp0};
(ii) reveal the leading  and intuitively simple role that the singularity of the long run conditional
error covariance matrix $\Omega_{00.x}$ plays in giving rise to
multicointegration, a feature of the model that may be unknown to the investigator; 
(iii) allow for both cointegration and multicointegration within the same
specification; 
and (iv) use a nonparametric formulation to provide a general setting for the analysis, for the form of the cointegrating and multicointegrating coefficients, and for practical work.

In a VAR framework Engsted and Johansen  (1999) show that  multicointegration,
as defined in Granger and Lee (1990) of a linear $I(1)$
process\footnote{ The summability condition in the specification \eqref{pp1} imposes a restriction on $C(z)$. This is because the matrix moving average power series $D(z)$ of a linear process generated 
by \eqref{pp0}-\eqref{pp1} does not have poles at $z=1$, which implies that,
when the system is written in the form $\left(y'_t,x'_t\right)'=\left(1-L\right)^{-1}C(L)\eta_t$,
it must be that
$\xi_1 C(1)=0$, where $\xi_1=(I_{m_0},  -A)$. Indeed, the upper block of $D(z)$
of such a system satisfies
$(D_{00}(z),  D_{0x}(z)) = \xi_1 (1-z)^{-1}C(z)=(1-z)^{-1}\xi_1 C(z)$ and does not
have poles at $z=1$ if and only if $\xi_1 C(1)=0$.
We thank a referee for this clarifying observation
and suggestions to improve the statement and the proof of
Proposition~\ref{prop:mcoint}.}
$\left(y'_t,x'_t\right)'=\left(1-L\right)^{-1}C(L)\eta_t$ where the roots of
$|C(z)|=0$ satisfy $|z|> 1$ or $z=1$,
occurs when $z=1$ is a root\footnote{The order $m$ of zeros of $C(z)$ at
$z=1$ or, equivalently, the order of poles of $C(z)^{-1}$ at $z=1$, is not restricted to $m=1$ and
is unknown. A unified treatment of the different representations of cointegrated systems for
known $m$ is given in a recent paper by Franchi and Paruolo (2019).}, so that $C(1)=\xi\epsilon'$ has reduced rank and $\xi'_{\perp}\dot
C(1)\epsilon_{\perp}$ is singular (explicit forms of the matrices $\{\xi,\epsilon\}$ and their orthogonal complements $\{\xi_{\perp}, \epsilon_{\perp}\}$ are given 
in the proof of the Proposition 1). This is the case when $\Omega$ is singular or more specifically in the present context when $\Omega_{00.x}$ is singular, as shown below. 
\begin{prop}\label{prop:mcoint}
A linear process $\left(y'_t,x'_t\right)'$ generated
by~\eqref{pp0}-\eqref{pp1} with $\Omega_{xx}>0$ is multicointegrated, i.e., $\Omega$ is singular, if and only if
it satisfies the multicointegration condition of Engsted and Johansen (1999). The rank of the multicointegrating relation equals
$m-rank(\Omega)=m_0-rank(\Omega_{0.xx})$.  
\end{prop}

In what follows data matrices are denoted by upper case letters without indexes, e.g., $Y'=[y_1,\ldots,y_T]$.  The OLS estimator $\widehat A= Y'X \left(X'X\right)^{-1}$ is consistent
at the rate at least $O(T)$.  The FM-OLS estimator (Phillips
and Hansen 1990) has the form $\widehat A^{+}=\left(\widehat
Y^{+'}X-T\widehat\Delta^{+}_{0x}\right) \left(X'X\right)^{-1}$ and employs corrections for endogeneity in the regressor $x_t$, leading to the transformed dependent variable $\widehat
y^{+}_t=y_t-\widehat\Omega_{0x}\widehat\Omega_{xx}^{-1}(x_t-x_{t-1})$ and a bias correction term involving 
$\widehat\Delta^{+}_{0x}
=\widehat\Delta_{0x}
-\widehat\Omega_{0x}\widehat\Omega_{xx}^{-1}\widehat\Delta_{xx}$,
which is constructed in the usual way using consistent nonparametric estimators
of submatrices of the long run and one sided long run quantities $\Omega$ and
$\Gamma^{+}$. Compared with OLS, the FM-OLS estimator removes asymptotic bias and increases efficiency by correcting
both the long run serial correlation in $u_t$ and endogeneity in $x_t$ caused by the
long run correlation between $u_{0t}$ and
$u_{xt}$. The properties of FM-OLS in general regressions as well as VARs are studied in Phillips (1995). Here we advance the analysis by allowing for the possibility of a singular conditional long
run variance matrix $\Omega_{00.x}$.  When $\Omega_{00.x}$ is singular, i.e. when modified $y_t$ is
cointegrated and in some direction the errors in the cointegrating equation
are $I(-1)$, the limit theory of the FM-OLS estimator is degenerate at the usual $O(T)$ rate and the faster convergence rate affects both estimation and inference. 
 
The paper makes the following contributions. First, we derive the new rates of
convergence and limit distribution of the FM-OLS estimator in the case of a null conditional long run
variance matrix. The new rate exceeds $O(T)$ and depends on the bandwidth used in estimating the long run covariance matrix quantities that are employed in making corrections for endogeneity
and serial correlation in FM-OLS.  The resulting limit distribution is no longer mixed normal and
depends on nuisance parameters. Similar properties hold in the direction of singularity
in the case of a singular long run variance matrix. Second, under certain conditions, the limit distribution of Wald statistics for testing restrictions on the cointegrating space and cointegrating parameters is $\chi^2$ and is invariant to the presence of singularity. Third, we show that when those restrictions fail, the Wald test is conservative. Monte Carlo simulations reveal that the empirical
level of the test can be far below the nominal $1\%$, $5\%$ and $10\%$ levels in singular and near
singular cases. 

As an application of our methods we analyze fiscal sustainability of the US
government over the period 1947-2019 by testing the null hypothesis that the
cointegration relationship between government revenue and expenditure has the parametric form $(1,-1)$. Multicointegration between government revenue and expenditure naturally arises if bounds are imposed on deviations 
of debt from revenue. We reject the null hypothesis and, as our theoretical results show,
this conclusion is not affected by the presence of multicointegration. The finding is
important for practical purposes, as a separate treatment of the
multicointegration case is not necessary (cf., Quintos, 1995, and
Berenguer-Rico and Carrion-i-Silvestre, 2011).

The paper is organized as follows.
In Section 2 we derive
the rates of convergence of elements of $\widehat A^{+}$ and establish its
limit distribution. After some preliminary observations we begin our
discussion with the null case where $\Omega_{00.x}=0$, then move on to a case of
a general singular matrix. The implications of singularity for hypothesis testing are discussed in Section
3. The finite sample properties of the FM-OLS and Wald test statistics are explored in Section~4. 
The application to government fiscal sustainability is considered in
Section~5.  Section 6 concludes. Proofs are given in the Appendix.

\section{Fully Modified OLS} 
Under the stated conditions the functional law $T^{-1/2}\sum_{t=1}^{[T\cdot]}u_t\to_d B(\cdot)\equiv BM(\Omega)$ holds for partial sums of $u_t$ (e.g., Phillips and Solo, 1992). Define the partition $B=(B'_0,B'_x)'$ into the first $m_0$ and the
final $m_x$ subvectors of the Brownian motion, conformably with $u_t$.  
Introducing the matrices
\begin{equation*}
\LL_{0.\Omega}=\left[\begin{array}{cc}
I_{m_0} &  -\Omega_{0x}\Omega_{xx}^{-1}
\end{array}%
\right],\quad
\LL_{\Omega}=\left[\begin{array}{cc}
I_{m_0} &  -\Omega_{0x}\Omega_{xx}^{-1} \\
0 & I_{m_x}%
\end{array}%
\right]
\end{equation*}
and the Brownian motion $B_{0.x}=\LL_{0.\Omega}B$, we have 
\begin{equation*}
\left[
\begin{array}{c}
B_{0.x} \\
B_{x}%
\end{array}%
\right]
=
\LL_{\Omega}\left[
\begin{array}{c}
B_{0} \\
B_x%
\end{array}%
\right]
=
BM(\LL_{\Omega}\Omega \LL'_{\Omega})
=
BM\left(\left[
\begin{array}{cc}
\Omega_{00.x}  & 0 \\
0 & \Omega_{xx}%
\end{array}%
\right]\right),
\end{equation*}
where $B_{0.x}\equiv BM\left(\Omega_{00.x}\right)$ is orthogonal to $B_x$.
Note that $\Omega_{00.x}$ is the long run variance of
$u_{0.x,t}=\LL_{0.\Omega}u_{t}=D_{0.x}(L)\eta_t$, where
$D_{0.x}(L)=\LL_{0.\Omega}D(L)$.
It is well known that the OLS estimator of $A$ in \eqref{pp0} is $O(T)$
consistent with a limit distribution that depends on the nuisance parameters
$\Omega$ and $\Gamma^{+}$, viz., 
\begin{align} \label{pp2}
T\left(\widehat A-A\right)\to_d&
\left(\int_0^1 d B_{0} B'_x + \Delta_{0x}\right) \left(\int_0^1 B_x B'_x\right)^{-1}
=\left(\int_0^1 d B_{0.x} B'_x\right) \left(\int_0^1 B_x B'_x\right)^{-1} \notag\\
&+\Omega_{0x}\Omega_{xx}^{-1}\left(\int_0^1 d B_{x} B'_x\right) \left(\int_0^1 B_x B'_x\right)^{-1}+
\Delta_{0x} \left(\int_0^1 B_x B'_x\right)^{-1}. 
\end{align} 
The last two terms of \eqref{pp2} are the endogeneity and serial correlation biases that
FM-OLS seeks to remove.

Suppose $\Omega$ and $\Gamma^{+}$ are estimated in the usual way (e.g., Priestley 1981; Hannan, 1970) as
\begin{align*}
\widehat\Omega=\sum_{j=-T+1}^{T-1}w(j/K)\widehat\Gamma_{\widehat u,\widehat u}(j)
\ \text{and} \ 
\widehat\Delta=\sum_{j=0}^{T-1}w(j/K)\widehat\Gamma_{\widehat u,\widehat u}(j),
\end{align*}
where $w(\cdot)$ is a kernel function, $K$ is a bandwidth parameter and the sample covariances are 
$\widehat\Gamma_{\widehat u,\widehat u}(j)=T^{-1}\sum_{1\le t,t+j\le T}\widehat u_{t+j} \widehat u'_{t},
\  \  \widehat u_{t}=(\widehat u'_{0t}, u'_{xt})'$,
where $\widehat u_{0t}=y_t-\widehat A x_t$. Similar to Phillips (1995), we consider the following kernels and bandwidth
rates. 

\begin{assumptiontwo}{K}{Kernel Condition}
For given $k\in(0,1)$, the bandwidth parameter $K$ has the rate $K\sim c_T T^k$ as $T\to\infty$,
where $c_T$ is slowly varying at infinity, i.e.\ $c_{xT}/c_{T}\to 1$ for $x>0$ and $T\to\infty$.
The kernel function $w(\cdot):\R\to[-1,1]$ is a twice continuously
differentiable even function  with 

\begin{enumerate}
\item[(a)] $w(0)=1, w'(0)=0, w''(0)\ne 0$, and
\item[(b)] $w(x)=0, |x|\ge 1$, with $\lim_{|x|\to 1}w(x)/(1-|x|)^2=const$.
\end{enumerate}
\end{assumptiontwo}
Parzen and Tukey--Hanning kernels satisfy Assumption K. 
The Bartlett--Priestley or quadratic spectral kernels do not satisfy Assumption
K but to use them in the following development these kernels need to satisfy  
\begin{enumerate}
\item[(b')] $w(x)=O(x^{-2})$, as $|x|\to \infty$
\end{enumerate}
and (a) with support $(-\infty, \infty)$.
Under Assumption K, with $0<k<1$, and any consistent estimator $\widehat A$ we have 
$\widehat\Gamma\to_p\Gamma,\ %
\widehat\Omega\to_p\Omega,\ %
\widehat\Delta\to_p\Gamma^{+}.$

\begin{prop}\label{prop:Rfull}
Under Assumption K with $0<k<1$,
\begin{align*}
T\left(\widehat A^{+} - A\right)\to_d
\left(\int_0^1 d B_{0.x} B'_x\right) \left(\int_0^1 B_x B'_x\right)^{-1}.
\end{align*} 
\end{prop}
For the nonsingular case this result appears in Corollary 4.3 in Phillips
(1995). The proof reveals that singularity does not alter the above convergence
but makes the limit distribution degenerate.
If $\Omega_{00.x}$ has full
rank, the rate of convergence of the FM-OLS estimator is determined by the rates of weak
convergence of the sample covariances and the rate of nonparametric estimation
of $\Omega$ and $\Gamma^{+}$ does not play any role.
We will show that in case $\Omega_{00.x}$ is singular, 
the rate of convergence of the FM-OLS estimator
along the null direction of $\Omega_{00.x}$ increases by $\delta(T)$, where
\begin{align*}
\delta(T)=\left\{
  \begin{array}{ll}
    T^{1+2k}, & k\in(0,1/4),\\
    T^{3/2}, & k\in[1/4,1/2],\\
    T^{2-k}, & k\in(1/2,1).
\end{array}
\right.
\end{align*}
The fastest rate of convergence of FM-OLS in the null direction is $T^{3/2}$ when bandwidth
expansion rate is $k\in[1/4,1/2]$.

For example, in the case where $\Omega_{00.x}=0$, we have
$T\left(\widehat A^{+} - A\right)\to_p 0$ and the precise rate of convergence of the FM-OLS estimator depends on the bandwidth parameter expansion rate $k$ in kernel estimation of the nonparametric components. Parameter dependencies may then be present in the resulting asymptotic theory, arising from first order terms in the limit behavior of the long run covariances that influence the asymptotics in this degenerate case.
In particular, when $\Omega_{00.x}=0$ the component $D_{0.x}(1)=0$ in the
Beveridge-Nelson decomposition of $u_{0.x,t}$. In this case, 
$u_{0.x,t}=D_{0.x}(1)\eta_t+\widetilde D_{0.x}(L)\eta_{t-1}-\widetilde D_{0.x}(L)\eta_t$
reduces to a first difference $u_{0.x,t}=-(e_t-e_{t-1})$, which is
$I(-1)$\footnote{The fact that $u_{0.x,t}$ is $I(-1)$ includes the possibility that in some or
all directions $u_{0.x,t}$ may be $I(-d)$, $d>1$, but that possibility does not affect the
convergence properties of the FM-OLS estimator as the next proposition shows.} with 
$e_t=\widetilde D_{0.x}(L)\eta_t=\sum_{j=0}^{\infty}\widetilde
D_{0.x,j}\eta_{t-j}$, with $\widetilde D_{0.x,j}=\LL_{0,\Omega}\sum_{t=j+1}^{\infty}D_t$, and $e_t$ has long run variance matrix $\Omega_{ee}=\widetilde D_{0.x}(1)\widetilde D_{0.x}(1)'$.

The next proposition establishes
convergence properties of FM-OLS for such time series.
It is particularly useful for the
case of a single cointegration relationship with $m_0=1$ (e.g., Phillips and Loretan, 1991), because singularity implies that the
conditional long run variance is zero. This reduction makes explicit the effect of
singularity on the convergence rates and serves as the basis of a general result.
\begin{prop}\label{prop:Rnull}
Suppose $\Omega_{00.x}=0$.  Under Assumption K with $0<k<1$,
\begin{align*}
\delta(T)\left(\widehat A^{+} - A\right)=O_p(1).
\end{align*} 
\end{prop}
\noindent As the proof of Proposition \ref{prop:Rnull} reveals, the limit distribution of the restandardized estimation error 
$\delta(T)(\widehat A^{+} - A)$ depends on nuisance
parameters associated with the nonparametric estimation of the long run covariance matrices. For kernel estimators, the limit depends on the covariance structure of the errors, on the bandwidth growth rate, and on the second derivative
of the kernel function. For illustration, consider the case when the bandwidth $K$ grows
slower than $T^{1/4}$, which includes the typical optimal bandwidth rate $T^{1/5}$ for long run
variance estimation. Under these conditions, we have the following limit theory. 
\begin{prop}\label{prop:Rnulldistr}
Suppose $\Omega_{00.x}=0$.
Under Assumption K with $k<1/4$,
\begin{align*}
T^{1+2k}\left(\widehat A^{+} - A\right)\to_d
w''(0)
\left(\Phi_0 +\Phi_{-\infty}\Omega^{-1}_{xx}\int_0^1 d B_x B'_x \right) 
\left(\int_0^1 B_x B'_x\right)^{-1},
\end{align*} 
where $\Phi_h=\sum_{j=h}^{\infty}\left(j+1/2\right)\E e_{t+j} u'_{xt}$.
\end{prop}
\noindent Unlike the corresponding limit theory in the nonsingular case
(Phillips, 1995; Phillips and Hansen, 1990; Phillips, 2014), the limit distribution of FM-OLS
now depends on the covariance structure of the errors $u_{xt}$ and $e_{t}$ and on the second derivative
of the kernel function. 

Next, consider a general case of singular $\Omega_{00.x}$ with rank $r<m_0$, so that
$\Omega$ has rank $r+m_x$. To isolate nondegenerate directions decompose
$\Omega_{00.x}=RR'$, where $R$ is an $m_0 \times r$ matrix of rank $r$.  Then $R'R$ has full rank, $R'u_{0.x,t}$ has full rank long run variance
matrix and Proposition \ref{prop:Rfull} applies  in this direction. In the
orthogonal direction\footnote{By the usual eigenvalue decomposition for symmetric matrices there is a set of orthonormal eigenvectors $\{q_i\}_{i=1}^m$ of $\Omega_{00.x}$, stacked as an orthogonal matrix $C$ and real
	eigenvalues $\lambda_i$ in decreasing order on diagonal matrix $\Lambda$,
	such that $\Omega_{00.x}=C\Lambda C'=\sum_{i=1}^{r}\lambda q_i q_i'$. In this
	notation, $C\Lambda^{1/2}=(R,0)$ and $R_{\perp}$ spans the space of
	eigenvectors corresponding to zero eigenvalues.}
 $R_{\perp}$, $R'_{\perp}u_{0.x,t}=-(e_t-e_{t-1})$ is $I(-1)$
 \footnote{This representation allows for $R'_{\perp}u_{0.x,t}$ being $I(-d)$ with $d>1$ in some directions.},
where
$e_t=R'_{\perp}\widetilde D_{0,x}(L)\eta_t$
has long run variance  $\Omega_{ee}=R'_{\perp}\widetilde D_{0,x}(1)\widetilde
D_{0,x}(1)'R_{\perp}$ and
Proposition \ref{prop:Rnull} applies, showing that elements in this direction
$R'_{\perp}A$ are estimated at a faster rate than $O(T)$.

We now state our first main result.
\begin{theorem}\label{thm:RR}
Suppose $\Omega_{00.x}=RR'$, where $R$ is an $m_0 \times r$ matrix with $rank(R)=r<m_0$.
Then under Assumption K
\begin{align*}
T \left(\widehat A^{+} - A\right)\to_d
\left(\int_0^1 d B_{0.x} B'_x\right) \left(\int_0^1 B_x B'_x\right)^{-1},
\end{align*} 
which is degenerate mixed normal. The limit distribution is not degenerate and has full rank in direction $R$ with 
\begin{align*}
T R'\left(\widehat A^{+} - A\right)\to_d
\left(\int_0^1 d B_{f.x} B'_x\right) \left(\int_0^1 B_x B'_x\right)^{-1},
\end{align*} 
where $B_{f.x}\equiv BM\left(\Omega_{ff.x}\right)$ and $\Omega_{ff.x}=R'RR'R$ is
the full rank $r \times r$ conditional long run variance matrix of
$R'U_{0.x}$.  
In the direction $R_{\perp}$ orthogonal to $R$ the convergence of $\widehat
A^{+}$ is at the faster rate $O(\delta(T))$ and $\delta(T) R'_{\perp}\left(\widehat A^{+} - A\right)=O_p(1).$
\end{theorem}

The FM-OLS estimator of a singular triangular system with multicointegration therefore has the following properties: (i) FM-OLS is consistent; (ii) the limit distribution is degenerate in the original coordinates; and (iii) rates of
convergence are $O(T)$ in nondegenerate directions and $O(\delta(T))$ in degenerate directions. In the degenerate direction
the limit distribution is of the type shown in Proposition~\ref{prop:Rnulldistr}. Singularity of the limit distribution means that care is needed when undertaking inference and these matters are considered in the next section. The situation is in some ways analogous to that of causality testing in cointegrated VAR regressions, as analyzed in Toda and Phillips (1993), and cointegrating regressions with cointegrated regressors, as analyzed in Phillips (1995). In the present case, it is necessary to analyze the directions of singularity of the long run covariance structure and the behavior of the estimates in these directions.

\section{Testing}
We consider the following hypothesis involving functions $\phi\in C_q^1(-\infty,\infty)$, the space of $q$ dimensional continuously differentiable functions 
$\mathcal{H}_0:\; \phi(vec(A))=0,$ where $vec\left(\cdot\right)$ is row vectorization.  
Suppose $\Omega_{00.x}=RR'$, where $R$ is an $m_0 \times r$ matrix of $rank(R)=r<m_0$, so that
$R'R$ is $r \times r$ nonsingular. Then under Assumption K
\begin{align*}
T \left(\widehat A^{+} - A\right)\to_d
\left(\int_0^1 d B_{0.x} B'_x\right) \left(\int_0^1 B_x B'_x\right)^{-1}\equiv \mathcal{MN}\left(0,\Omega_{00.x}\otimes\left(\int_0^1 B_x B'_x\right)^{-1}\right).
\end{align*} 
The limit distribution is mixed normal ($\mathcal{MN}$) and standard inference methods
can be applied.  The usual Wald statistic for testing $\mathcal{H}_0$ is 
\begin{align*}
W=\phi\left(\widehat a^{+}\right)'
\left\{\Phi\left(\widehat a^{+}\right)
\left(\widehat\Omega_{00.x}\otimes\left(X'X\right)^{-1}\right)
\Phi\left(\widehat a^{+}\right)'\right\}^{-1}
\phi\left(\widehat a^{+}\right),
\end{align*}
where $\widehat a^{+}=vec\left(\widehat A^{+}\right)$, $\Phi(a)=\partial \phi(a)/\partial a'$ and $a=vec\left(A\right)$ is row vectorization. Suppose that the following rank condition holds
\begin{align}
rank \left\{\Phi\left(a\right)
\left(\Omega_{00.x}\otimes\left(\int_0^1B_x B'_x\right)^{-1}\right)
\Phi\left(a\right)'\right\}=q. \label{tt1}
\end{align}
Under Assumption K, $W\to_d\chi^2_q$. So, under the rank condition \eqref{tt1}, the limit
distribution of the Wald statistics is invariant to the presence of
singularity.

\subsection{Violation of the rank condition} 

Consider the linear hypothesis 
\begin{align*}
\mathcal{H}_0:\; Q vec(A)=r_0,
\end{align*}
with restriction $q\times m_0 m_x$ matrix $Q=R_1\otimes R_2$ of rank $q=q_1q_2$ and component matrices $R_1$, $m_0 \times q_1$, of rank $q_1$ and $R_2$, $m_0 \times q_2$, of rank $q_2$. Then $\mathcal{H}_0$ has a tensor form with matrix representation
$\mathcal{H}_0:\; R_1 A R_2=R_3, \; \textrm{with} \; vec\ R_3=r_0$, and 
\begin{align*}
Q
\left(\Omega_{00.x}\otimes\left(\int_0^1 B_x B'_x\right)^{-1}\right)
Q'
=
R_1\Omega_{00.x}R'_1 \otimes R'_2 \left(\int_0^1 B_x B'_x \right)R_2.
\end{align*}
If the rank of $ R_1\Omega_{00.x}R'_1=R_1 RR' R'_1$ is $\tilde q_1<q_1$, then
the rank condition \eqref{tt1} fails as $\tilde q_1 q_2< q_1 q_2=q$.
This is the case where some of the restrictions isolate
directions in which FM-OLS is hyperconsistent with  rate exceeding $O(T)$. 
The distribution of the Wald test statistic is then nonstandard and depends on nuisance
parameters. In general, failure of estimator mixed normality in the direction of faster convergence produces a non
chi-squared limit in the Wald statistic as the faster convergence of the
estimator is balanced in the Wald statistic weighting. A related phenomenon
arises in Toda and Phillips (1993), who describe situations where Wald tests
of Granger causality in cointegrated VAR systems do not follow asymptotically chi-squared distributions. For another example, see Phillips (2016), where singularity in the signal matrix leads to nonstandard inference.

To illustrate the consequences of singularity consider testing
$\mathcal{H}_0: A=A^0.$
Then $Q=I_{m_0 m_x}, r_0=vec(A^0), R_1=I_{m_0}, R_2=I_{m_x}, R_3=A^0$ and
their ranks are $q=m_0 m_x$, $q_1=m_0$ and $q_2=m_x$.
The Wald test statistic then simplifies to
\begin{align*}
W_I&=vec\left(\widehat A^{+}-A^{0}\right)'
\left(\widehat\Omega_{00.x}\otimes\left(X'X\right)^{-1}\right)^{-1}
vec\left(\widehat A^{+}-A^{0}\right)\\
&=
tr\left\{
\left(X'X\right)
\left(\widehat A^{+}-A^{0}\right)'
\widehat\Omega_{00.x}^{-1}
\left(\widehat A^{+}-A^{0}\right)
\right\}.
\end{align*}
The notational change to $W_I$ emphasizes that the following analysis 
only considers the full dimensional restriction structures above.
The rank of 
$\Omega_{00.x} \otimes \int_0^1 B_x B'_x$
equals the rank of the conditional long run variance times $m_x$, i.e. the null hypothesis restrictions isolate `all directions' and the rank condition is satisfied if and only if the conditional long run variance matrix is nonsingular.
If the conditional long run variance is nonsingular, the rank condition
holds and $W_I\to \chi^2_{q}$.

Singularity alters the rate of convergence and the limit  of $\widehat\Omega_{00.x}$,
which is used in the construction of the Wald test statistics. We proceed to
derive the rate of convergence for this quantity. 
As the proof of the following proposition reveals,
the rate of convergence is $T^{2k}$ if $k<1/3$ and $T^{1-k}$ if $k\ge 1/3$.
As in the case of FM-OLS estimation, 
the limit distribution of $\widehat\Omega_{00.x}$
depends on nuisance parameters and on the implementation of the nonparametric estimates of the long run
covariance matrices. As an illustration, consider a case when the bandwidth $K$ grows
slower than $T^{1/3}$, which includes the usual optimal bandwidth rate $T^{1/5}$ for long run
variance estimation. 
\begin{prop}\label{prop:RnullDenomDistr}
Suppose $\Omega_{00.x}=0$ and Assumption K holds with $0<k<1/3$.Then

\begin{enumerate}[(a)]
\item
$
T^{2k}\widehat\Omega_{00.x}
\to_p
-w''(0)\Omega_{ee}.
$
\item 
If $\Omega_{ee}$ is nonsingular, then $W_I=O_p(T^{-2k})$ if $k\in(0,1/4]$ and
$W_I=O_p(T^{2k-1})$ if $k\in(1/4,1/3)$.
\end{enumerate}
\end{prop}

Nonsingularity of $\Omega_{ee}$ means that there is no further level
of cointegration (system~\eqref{pp0} is singular of first order, see Park 1992)
and guarantees that the limit in Part~(a) is nondegenerate, so
that the rate of convergence of $\widehat\Omega_{00.x}$ is sharp. In the more general case where $\Omega_{00.x}$ is positive semi-definite but not a null matrix we have the following result.

\begin{theorem}\label{thm:Wald}
Suppose $\Omega_{00.x}=RR'$, where $R$ is an $m_0 \times r$ matrix with $rank(R)=r<m_0$,
$\Omega_{ee}$ is nonsingular,
and Assumption K holds with $0<k<1/3$, 
then under the null $W_I\to_d \chi^2_{r m_x}$.
\end{theorem}
The proof of the above result reveals that the limit distribution of the Wald test
statistic involves the sum of two major components. The first component arises from the limit in the
nonsingular direction, which is $\chi^2_{r m_x}$, and the second involves the limit in the
direction where the conditional long run variance matrix is zero, which is nonstandard,
depends on nuisance parameters and decays at the speed established in Proposition \ref{prop:RnullDenomDistr} above.
Therefore, for $k<1/3$ the $\chi^2_{r m_x}$
limit distribution of the $W_I$ statistic has thinner tails than 
the distribution of $\chi^2_{m_0 m_x}$, so that tests based on the usual degrees of freedom are asymptotically conservative.

\section{Finite sample performance}
The following analysis of finite sample performance is based on $10,000$
simulations with different sample sizes.\footnote{Statistical computing in
this paper uses R version 3.4.4.} 
The long run variances are estimated using the Parzen kernel
\begin{equation*}
w(x)=\left\{\begin{array}{cc}
1-6x^2+6|x|^3,& \quad -1/2\le x \le 1/2,\\
2(1-|x|)^3,& \quad 1/2\le |x| \le 1,\\
0,& \quad 1\le |x|,
\end{array}%
\right.
\end{equation*}
and bandwidth is set to $K=T^{1/4}$, if not specified otherwise.
The data generating process (DGP) has the form \eqref{pp0}
with scalar cointegrating coefficient $A=2$ and with a combined error vector $u_t = (u_{0t}, u'_{xt})'$ that follows the linear process
\begin{equation*}
u_t = \eta_t + D_1\eta_{t-1}, \quad\text{with} \quad \eta_t\sim iid\, \mathcal{N}(0,\Sigma).
\end{equation*}
We consider estimation of $A$ and hypothesis testing for the null $\mathcal{H}_0:\ A=2$.
We look at two classes of bivariate ($m=2$) DGPs.

\begin{enumerate}[DGP1]
\item
For parameter choices $p\in \{0.0, -0.1, \ldots, -1.0\}$ define
\begin{equation*}
D_1=\left[
\begin{array}{cc}
p & 0 \\
0 & 0%
\end{array}%
\right],
\quad\Sigma=I_2.
\end{equation*}
\item
For parameter choices $p\in \{0.8, 5.2\}$ define
\begin{equation*}
D_1=\left[
\begin{array}{cc}
0.3 & 0.4 \\
p & 0.6%
\end{array}%
\right],
\quad
\Sigma=\left[
\begin{array}{cc}
1 & 0.5 \\
0.5 & 1%
\end{array}%
\right].
\end{equation*}
\end{enumerate}

\begin{figure}[!ht]
\centering
\subfigure[Bias, $T=50$] {
\includegraphics[width=0.47\textwidth]{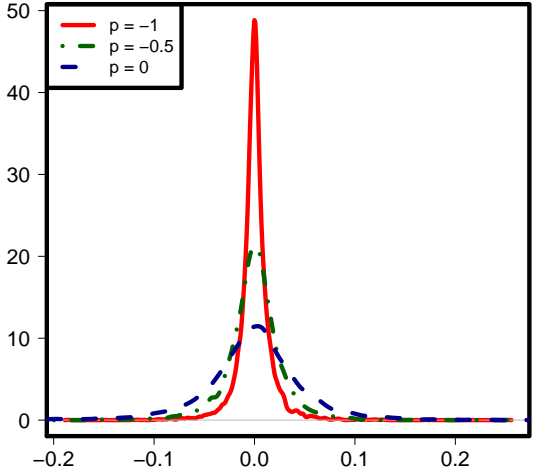}
}
\subfigure[$t$-statistics, $T=50$] {
\includegraphics[width=0.47\textwidth]{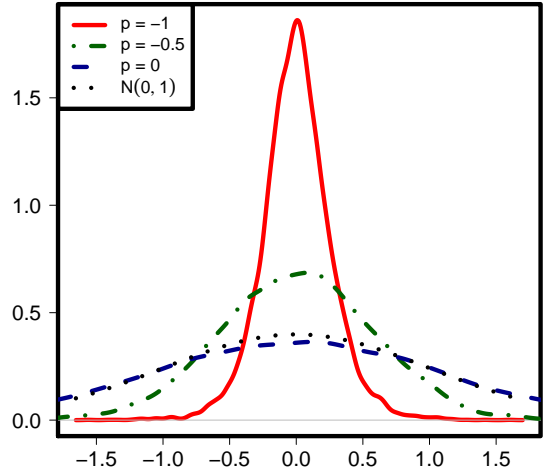}
}
\subfigure[Bias, $T=100$] {
\includegraphics[width=0.47\textwidth]{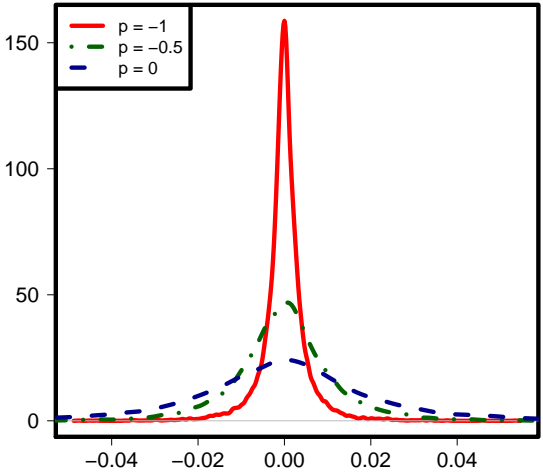}
}
\subfigure[$t$-statistics, $T=100$] {
\includegraphics[width=0.47\textwidth]{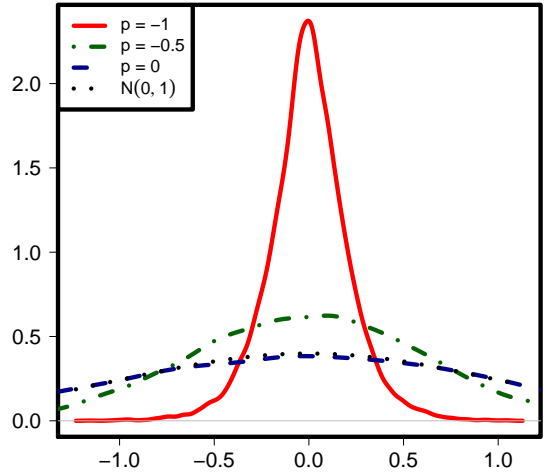}
} 
\caption{Kernel estimates of the densities of the FM-OLS estimation error $\widehat A^{+}-A$ and $t$-statistic $t_{\widehat A^{+}}$ for sample sizes
$T=50$ and $T=100$ and DGP1 with parameters $p\in\{0.0,-0.5,-1.0\}$.
Multicointegration occurs if $p=-1.0$.} 
\label{fig:1}
\end{figure}
\begin{figure}[!ht]
\centering
\subfigure[Bias, $T=50$] {
\includegraphics[width=0.47\textwidth]{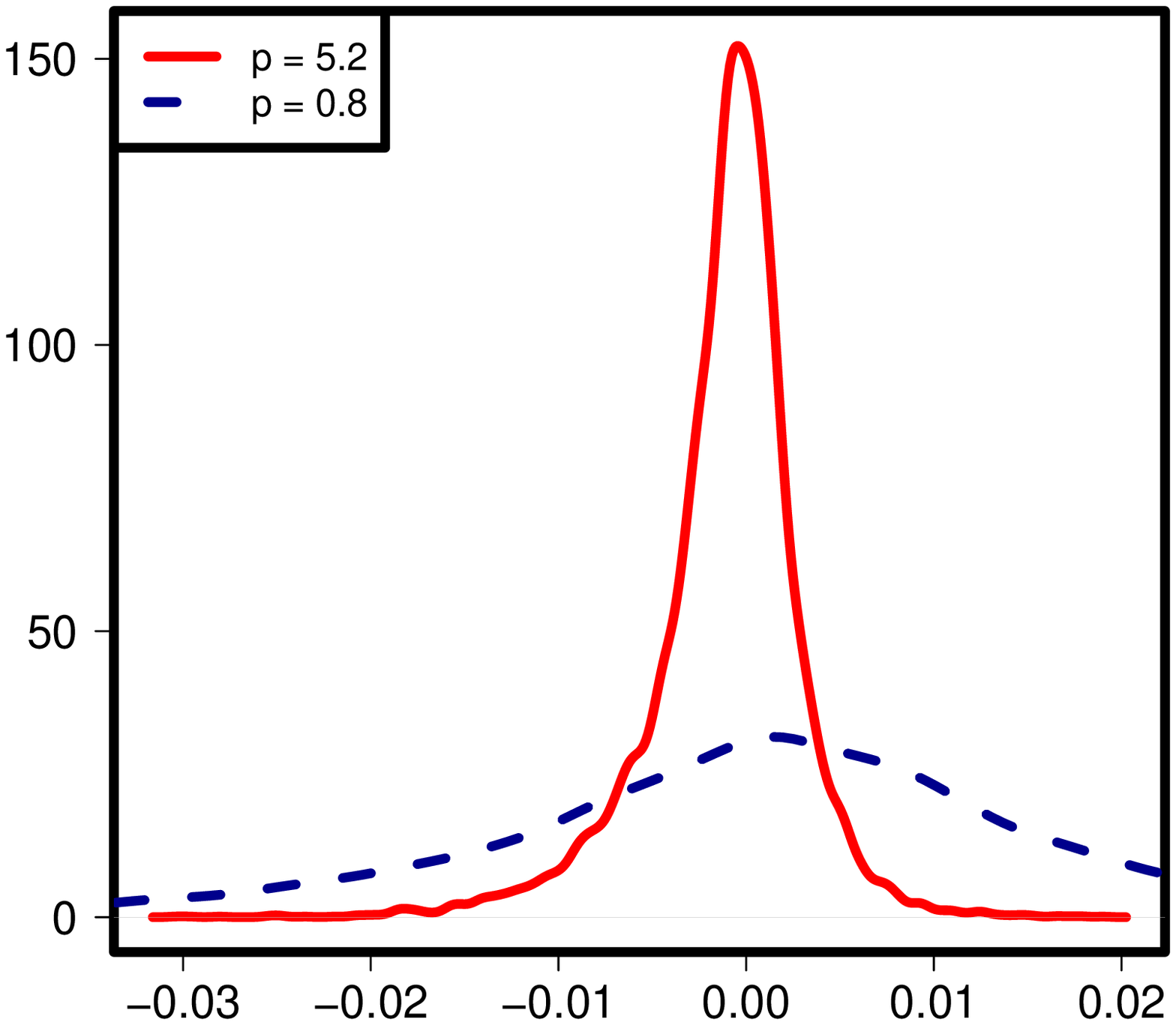}
}
\subfigure[$t$-statistic, $T=50$] {
\includegraphics[width=0.47\textwidth]{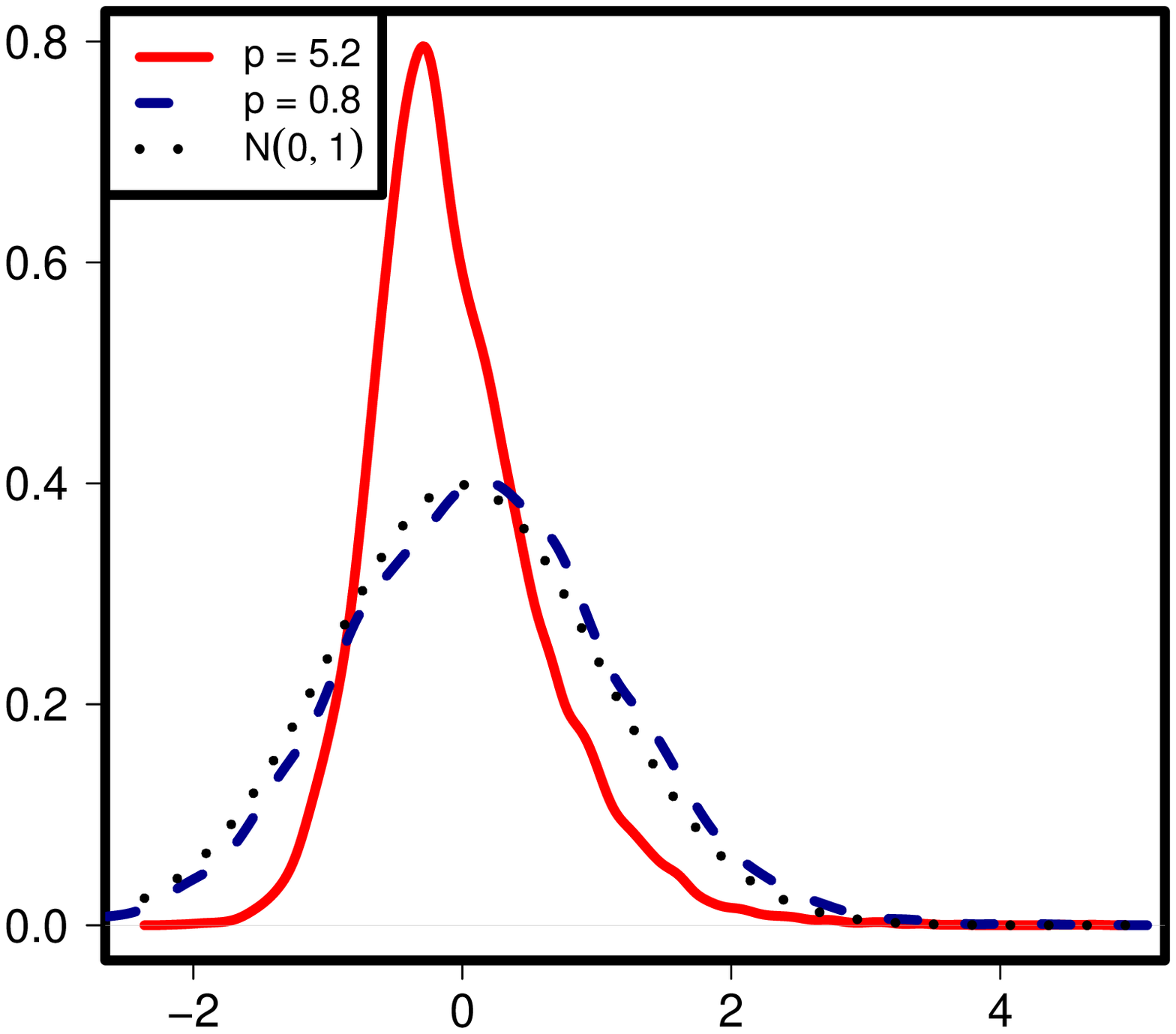}
}
\subfigure[Bias, $T=100$] {
\includegraphics[width=0.47\textwidth]{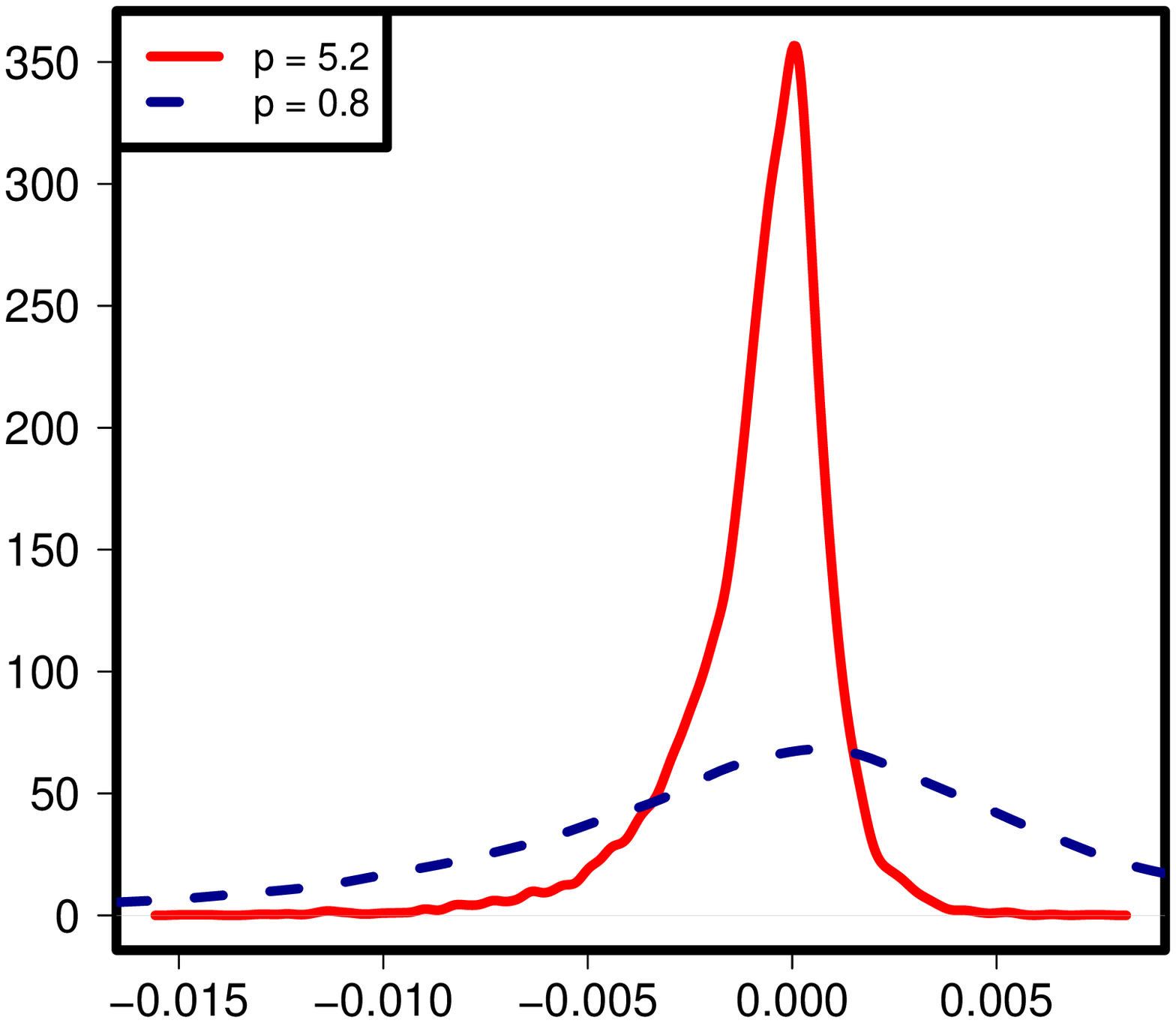}
}
\subfigure[$t$-statistic, $T=100$] {
\includegraphics[width=0.47\textwidth]{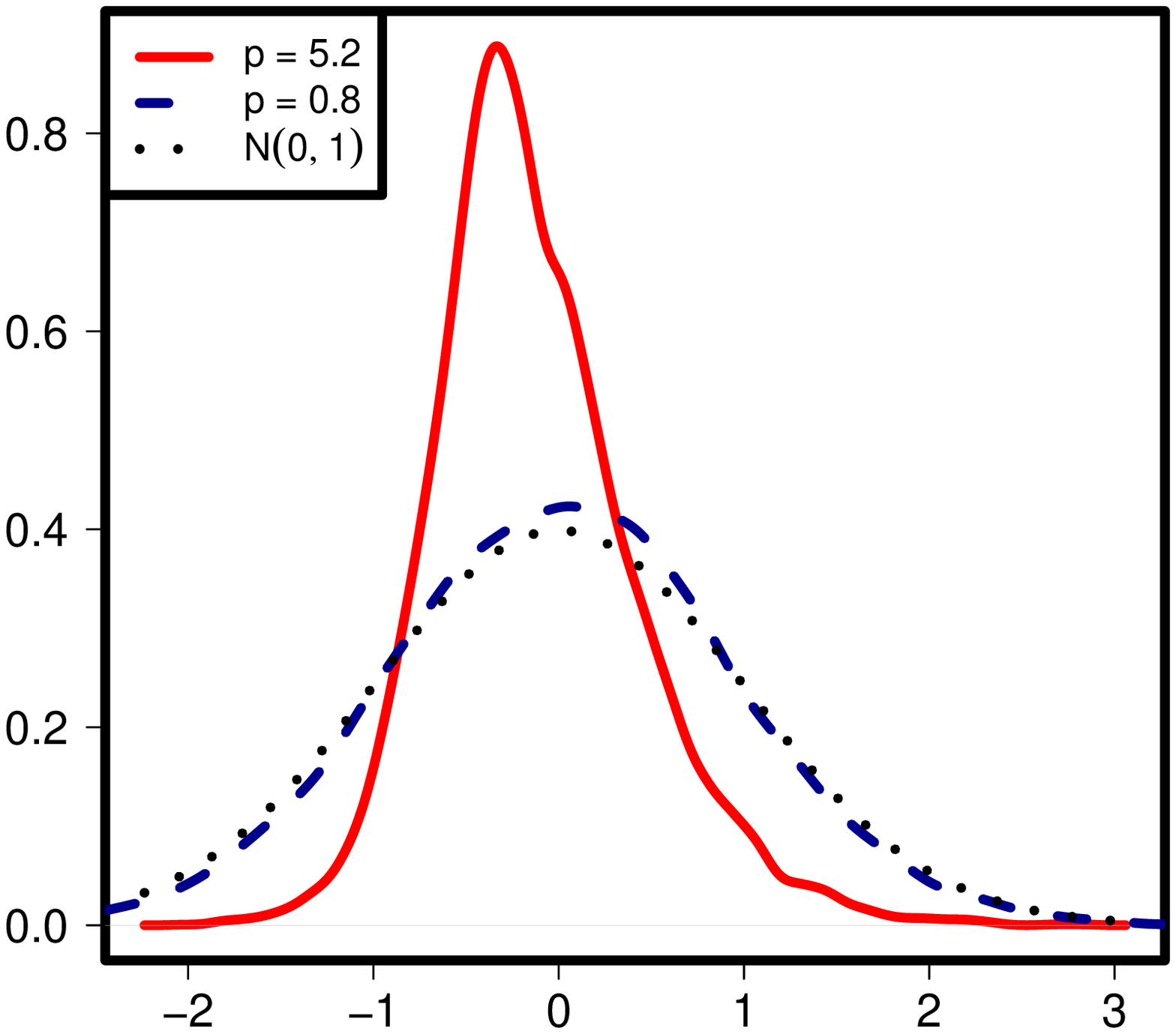}
} 
\caption{Kernel estimates of the densities of the FM-OLS estimation error $\widehat A^{+}-A$ and $t$-statistic $t_{\widehat A^{+}}$ for sample sizes $T=50$ and $T=100$ and DGP2 with parameters $p=5.2$  and $p=0.8$.
Multicointegration occurs if $p=5.2$.} 
\label{fig:2}
\end{figure}
\noindent The finite sample performance for DGP1 is shown in Figure~1 and Table~1;
results for DGP2 are given in Figure~2 and Table~2.

\noindent \textbf{Discussion of results for DGP1.}

\vspace{0.5mm}
\noindent When $p=-1$ we have $\Omega_{00.x}=0$ and a singular system. The limit theory generalizes results on estimation and testing to this case. With matrix $D_1$ diagonal, the errors $u_{0t}$ and $u_{xt}$ are independent
and the effect of singularity in the long run variance can be studied separately from the effect of long
run dependence. When $p=0$, the long run variance is the identity $I_2$ and the conditional
long run variance is $1$, giving a standard nonsingular case. For values of $p$
between $0$ and $-1$ the system is still in the nonsingular case but in finite samples for smaller values of $p$
the limit theory for the singular case may lead to better approximations than the nonsingular case and simulations help to guide this assessment.

In Figure~\ref{fig:1}, Panel~(a), the densities of the bias 
$\widehat A^{+}-A$ are shown for sample size $T=50$. We compare
the densities in the singular case ($p=-1$) with two nonsingular cases ($p=0$ and
$p=-0.5$). The figure shows that the bias in the singular case is much smaller than the
bias in the nonsingular cases. A more pronounced effect is observed for
$T=100$ in Panel~(c) confirming the higher convergence rates established for FM-OLS under
singularity.

We use the $t$-statistic 
$t_{\widehat A^{+}}=(\widehat
A^{+}-A^0)/\{\widehat\Omega_{00.x}/\sum_{t=1}^T x_t^2\}^{1/2}$
for testing the hypothesis $\mathcal{H}_0:\, A=2$. 
In Figure~\ref{fig:1}, Panel~(b) the densities of the $t$-statistic are shown
for sample size $T=50$. We compare
the densities in the singular case ($p=-1$) with two nonsingular cases ($p=0$ and
$p=-0.5$).  Theory predicts that in nonsingular cases the test statistics is asymptotically standard
normal, whose
density is also plotted. This approximation is quite accurate for $p=0$.
However, the
density of the test statistic for $p=-0.5$ has thinner tails, so that
the test based on standard normal approximation  is conservative as our theory
predicts for singular case. Results for the sample size $T=100$ are plotted in Panel~(d), and it is evident that the test statistic for $p=-0.5$ still has thin tails.

\begin{table}[!ht]
\begin{center}
\caption{The mean and standard deviation of the bias and $t$-statistics, the rejection rates for the
nominal $0.01$, $0.05$, and $0.10$ levels of the $t$-statistics based on the
FM-OLS and the mean and standard deviation of the bias of the OLS estimator are shown for
DGP1 for various values of $p$, sample sizes and bandwidths.
Multicointegration occurs when $p=-1$.} \vspace{2mm}
\label{t:sim1}
\setlength{\tabcolsep}{2.5pt}
%\footnotesize
%\scriptsize
%\tiny
\renewcommand{\arraystretch}{0.8}
\begin{tabular}{rrrrrrrrrrrrr}
& $T$ & $p$ & $K$ & Bias-OLS & SD-OLS & Bias & SD & t-Bias & t-SD & 0.10 & 0.05 & 0.01 \\ 
\hline
% latex table generated in R 3.5.1 by xtable 1.8-4 package
% Wed Jul 29 01:33:23 2020
% & T & p & K & Bias-OLS & SD-OLS & Bias & SD & t-Bias & t-SD & 0.90 & 0.95 & 0.99 \\ 
% \hline
1 & 50 & -1.0 & 2 & 0.0001 & 0.0238 & 0.0003 & 0.0175 & 0.0044 & 0.2598 & 0.000 & 0.000 & 0.000 \\ 
  2 & 100 & -1.0 & 3 & -0.0001 & 0.0088 & -0.0000 & 0.0055 & -0.0023 & 0.2124 & 0.000 & 0.000 & 0.000 \\ 
  3 & 50 & -0.9 & 2 & 0.0000 & 0.0231 & 0.0003 & 0.0173 & 0.0047 & 0.2752 & 0.000 & 0.000 & 0.000 \\ 
  4 & 100 & -0.9 & 3 & -0.0001 & 0.0087 & -0.0001 & 0.0057 & -0.0031 & 0.2379 & 0.000 & 0.000 & 0.000 \\ 
  5 & 50 & -0.8 & 2 & -0.0000 & 0.0234 & 0.0002 & 0.0186 & 0.0050 & 0.3214 & 0.000 & 0.000 & 0.000 \\ 
  6 & 100 & -0.8 & 3 & -0.0001 & 0.0091 & -0.0001 & 0.0068 & -0.0041 & 0.3108 & 0.000 & 0.000 & 0.000 \\ 
  7 & 50 & -0.7 & 2 & -0.0001 & 0.0247 & 0.0001 & 0.0210 & 0.0053 & 0.3937 & 0.000 & 0.000 & 0.000 \\ 
  8 & 100 & -0.7 & 3 & -0.0001 & 0.0101 & -0.0001 & 0.0085 & -0.0050 & 0.4112 & 0.000 & 0.000 & 0.000 \\ 
  9 & 50 & -0.6 & 2 & -0.0001 & 0.0267 & 0.0001 & 0.0242 & 0.0056 & 0.4846 & 0.001 & 0.000 & 0.000 \\ 
  10 & 100 & -0.6 & 3 & -0.0001 & 0.0115 & -0.0001 & 0.0105 & -0.0059 & 0.5232 & 0.002 & 0.000 & 0.000 \\ 
  11 & 50 & -0.5 & 2 & -0.0002 & 0.0294 & 0.0000 & 0.0280 & 0.0060 & 0.5875 & 0.007 & 0.002 & 0.000 \\ 
  12 & 100 & -0.5 & 3 & -0.0001 & 0.0132 & -0.0001 & 0.0126 & -0.0068 & 0.6363 & 0.010 & 0.003 & 0.000 \\ 
  13 & 50 & -0.4 & 2 & -0.0003 & 0.0326 & -0.0001 & 0.0321 & 0.0063 & 0.6967 & 0.019 & 0.007 & 0.000 \\ 
  14 & 100 & -0.4 & 3 & -0.0001 & 0.0151 & -0.0001 & 0.0149 & -0.0077 & 0.7433 & 0.027 & 0.009 & 0.001 \\ 
  15 & 50 & -0.3 & 2 & -0.0003 & 0.0361 & -0.0001 & 0.0364 & 0.0066 & 0.8073 & 0.041 & 0.017 & 0.003 \\ 
  16 & 100 & -0.3 & 3 & -0.0001 & 0.0172 & -0.0001 & 0.0171 & -0.0084 & 0.8392 & 0.050 & 0.019 & 0.002 \\ 
  17 & 50 & -0.2 & 2 & -0.0004 & 0.0399 & -0.0002 & 0.0409 & 0.0069 & 0.9146 & 0.071 & 0.031 & 0.007 \\ 
  18 & 100 & -0.2 & 3 & -0.0001 & 0.0192 & -0.0001 & 0.0195 & -0.0090 & 0.9215 & 0.075 & 0.034 & 0.006 \\ 
  19 & 50 & -0.1 & 2 & -0.0005 & 0.0439 & -0.0003 & 0.0455 & 0.0071 & 1.0146 & 0.106 & 0.053 & 0.013 \\ 
  20 & 100 & -0.1 & 3 & -0.0001 & 0.0214 & -0.0001 & 0.0218 & -0.0094 & 0.9894 & 0.097 & 0.049 & 0.009 \\ 
  21 & 50 & 0.0 & 2 & -0.0005 & 0.0481 & -0.0003 & 0.0502 & 0.0073 & 1.1043 & 0.134 & 0.073 & 0.021 \\ 
  22 & 100 & 0.0 & 3 & -0.0002 & 0.0236 & -0.0001 & 0.0242 & -0.0098 & 1.0438 & 0.116 & 0.061 & 0.014 \\ 
   \hline

\end{tabular}
\end{center}
\end{table}
\begin{table}[!ht]
\begin{center}
\caption{The mean and standard deviation of the bias and $t$-statistics, the rejection rates for the
nominal $0.01$, $0.05$, and $0.10$ levels of the $t$-statistics based on 
FM-OLS and the mean and standard deviation of the bias of the OLS estimator are shown for
DGP2 for various values of $p$, sample sizes and bandwidths.
Multicointegration occurs when $p=5.2$.} \vspace{2mm}
\label{t:sim2}
\setlength{\tabcolsep}{2.5pt}
%\footnotesize
%\scriptsize
%\tiny
\renewcommand{\arraystretch}{0.8}
\begin{tabular}{rrrrrrrrrrrrr}
& $T$ & $p$ & $K$ & Bias-OLS & SD-OLS & Bias & SD & t-Bias & t-SD & 0.10 & 0.05 & 0.01 \\ 
\hline
% latex table generated in R 3.5.1 by xtable 1.8-4 package
% Wed Jul 29 01:35:18 2020
% & T & p & K & Bias-OLS & SD-OLS & Bias & SD & t-Bias & t-SD & 0.90 & 0.95 & 0.99 \\ 
%  \hline
1 & 50 & 0.8 & 3 & 0.0207 & 0.0223 & 0.0013 & 0.0184 & 0.1368 & 1.0091 & 0.102 & 0.054 & 0.014 \\ 
  2 & 50 & 5.2 & 3 & 0.0026 & 0.0054 & -0.0011 & 0.0040 & -0.0442 & 0.6474 & 0.018 & 0.009 & 0.003 \\ 
  3 & 50 & 0.8 & 5 & 0.0207 & 0.0223 & 0.0039 & 0.0189 & 0.2638 & 1.0806 & 0.135 & 0.078 & 0.023 \\ 
  4 & 50 & 5.2 & 5 & 0.0026 & 0.0054 & 0.0003 & 0.0029 & 0.1242 & 0.6440 & 0.026 & 0.014 & 0.004 \\ 
  5 & 50 & 0.8 & 7 & 0.0207 & 0.0223 & 0.0056 & 0.0197 & 0.3659 & 1.1444 & 0.161 & 0.100 & 0.035 \\ 
  6 & 50 & 5.2 & 7 & 0.0026 & 0.0054 & 0.0008 & 0.0029 & 0.2425 & 0.7146 & 0.044 & 0.025 & 0.008 \\ 
  7 & 50 & 0.8 & 10 & 0.0207 & 0.0223 & 0.0076 & 0.0206 & 0.4959 & 1.2321 & 0.195 & 0.130 & 0.055 \\ 
  8 & 50 & 5.2 & 10 & 0.0026 & 0.0054 & 0.0012 & 0.0031 & 0.3657 & 0.8392 & 0.074 & 0.045 & 0.018 \\ 
  9 & 100 & 0.8 & 3 & 0.0104 & 0.0111 & -0.0003 & 0.0086 & 0.0146 & 0.9389 & 0.081 & 0.038 & 0.008 \\ 
  10 & 100 & 5.2 & 3 & 0.0013 & 0.0027 & -0.0008 & 0.0019 & -0.1256 & 0.5461 & 0.007 & 0.003 & 0.000 \\ 
  11 & 100 & 0.8 & 5 & 0.0104 & 0.0111 & 0.0007 & 0.0087 & 0.1036 & 0.9928 & 0.098 & 0.052 & 0.013 \\ 
  12 & 100 & 5.2 & 5 & 0.0013 & 0.0027 & -0.0002 & 0.0011 & 0.0058 & 0.4907 & 0.007 & 0.003 & 0.000 \\ 
  13 & 100 & 0.8 & 7 & 0.0104 & 0.0111 & 0.0014 & 0.0089 & 0.1700 & 1.0314 & 0.114 & 0.060 & 0.017 \\ 
  14 & 100 & 5.2 & 7 & 0.0013 & 0.0027 & 0.0001 & 0.0010 & 0.0977 & 0.5234 & 0.013 & 0.005 & 0.001 \\ 
  15 & 100 & 0.8 & 10 & 0.0104 & 0.0111 & 0.0020 & 0.0092 & 0.2500 & 1.0780 & 0.134 & 0.078 & 0.024 \\ 
  16 & 100 & 5.2 & 10 & 0.0013 & 0.0027 & 0.0003 & 0.0010 & 0.1937 & 0.5980 & 0.024 & 0.012 & 0.003 \\ 
   \hline

\end{tabular}
\end{center}
\end{table}

Further simulation results are in 
Table~\ref{t:sim1}. 
We vary sample sizes from $50$ to $100$ and the bandwidth  is set to $K=T^{1/4}$.
The bias is zero up to the 3d digit in all cases. When the sample size increases from $T=50$ to $T=100$ the precision of the FM-OLS measured by the standard deviation of the bias term increases by
$0.0175/0.0055=3.18$ in the singular case  and by $0.0502/0.0242=2.07$ in nonsingular case $p=0$ corroborating the hyperconsistency of FM-OLS in the singular case and superconsistency in nonsingular case.
When we compare FM-OLS with OLS, the former is more precise in the singular and
near-singular cases, while in the nonsingular case both estimators are
comparable as removing second order bias effects that do not exist under DGP1 in OLS does not give an
advantage to FM-OLS.

Interestingly, the rejection rates show that for $T=50$ the test is conservative in all
cases except $p=0$ and $p=-0.1$. Even when the sample size is raised to the high
level $T=10,000$ (unreported here)
the test is still conservative for values of $p=0.5$ and below. Thus, the phenomenon described
in this paper extends far beyond the pure singular case.

\noindent \textbf{Discussion of results for DGP2.}

\vspace{1.5mm}
Since Phillips and Loretan (1991) numerous simulation studies have considered
this DGP with different degrees of endogeneity, by varying  values for  $p$
between $-0.8$ and $0.8$. We consider $p=5.2$, for which $\Omega_{00.x}=0$ and
compare these results to the nearest nonsingular case with $p=0.8$.

In Figure~\ref{fig:2} we again observe thinner tails in the bias and $t$-statistic densities in
the singular case compared to the nonsingular case, as well as thinner tails in the density of the
$t$-statistic compared to the standard normal density already for $T=50$. Unlike
the situation with DGP1, second order biases are evident due to long run
covariance between $e_t$ and $u_{xt}$ as predicted by theory and indicated in the proofs in the Appendix.

 Table~\ref{t:sim2} confirms the above findings and allows analysis of the effect of bandwidth choice. Unlike  previous simulations, the bandwidth is here fixed for $K$ to a range between $3$ and
$10$. This range includes many typical rules for bandwidth choice for the sample
sizes considered. We see that test statistics for the singular case
control size even for bandwidth choices for which the nonsingular case results in
over-rejection ($T=50$, $K=10$). In all other cases singularity in the model results in a
conservative test for all bandwidth choices under consideration.

\section{Evaluating Fiscal Sustainability}

Soaring government debt in many countries calls for better economic understanding of
fiscal sustainability for which improved methods of econometric analysis may be helpful given the presence of nonstationarity and endogeneities in the relevant data. Econometric analysis of sustainability has a long tradition, going back to early work by Hamilton and Flavin (1986) who suggested to test stationarity of the discounted debt. Hakkio and Rush (1991), Huag (1991), Trehan and Walsh (1991), and Quintos (1995) were among the
first to test cointegration between revenues and expenditures. Quintos (1995) calls sustainability `strong' when revenues and expenditures cointegrate with the explicit coefficients $(1,-1)$ and tests the later using FM-OLS based $t$-statistics. 
A recent discussion of other approaches to evaluate fiscal sustainability is given in the chapter by D'Erasmo, Mendoza and Zhang (2016) in the Handbook of Macroeconomics.

Two remarks concerning the cointegration approach are relevant to our following analysis.
First, cointegration between revenues and expenditures is only a sufficient condition for an intertemporal budget
constraint (IBC) to hold and there are many other data generating processes
consistent with IBC. This means that rejecting cointegration does not imply that
IBC does not hold.  Following Bohn (2007), consider
\begin{align*}
B_t = B_{t-1} + G_t - R_t = G^0_t - R_t + (1+r_t)B_{t-1},\quad\text{Budget Identity (BI)},
\end{align*}
where $B_t$ is government debt, $R_t$ is government revenue, $r_t$ is the interest rate,
which is assumed to be stationary with mean $r>0$, 
$G_t$ is government expenditure, $G^0_t$ is government expenditure excluding
interest on debt, and $G^a_t=G^0_t+(r_t-r)B_{t-1}$ is adjusted expenditure.
These variables can be defined in nominal or real terms, possibly deflated by
GDP or population. For example, Quintos (1995) constructed real variables by
deflating nominal variables by the GNP price deflator and by population. 
BI implies
\begin{align*}
B_t = \frac{1}{1+r} \E_t G^0_t + (1+r_t)B_{t-1},
\quad\text{Difference Equation (DE)},
\end{align*}
which together with 
\begin{align*}
B_t = \lim_{j\to\infty}\frac{1}{(1+r)^j} \E_t B_{t+j} = 0,\quad (m.s.),
\quad\text{Transversality Condition (TC)},
\end{align*}
where the limit is in the mean square sense, implies
\begin{align*}
B_t = \sum_{j=1}^{\infty}\frac{1}{(1+r)^j} \E_t (R_{t+j} - G^a_{t+j}),
\quad\text{Intertemporal Budget Constraint (IBC)}.
\end{align*}
IBC holds when the debt matches the expected present discounted value of the future
surplus, a desirable requirement for sustainability.  Bohn (2007) shows that if $B_t\sim I(m)$ for some finite $m\ge 0$, then $B_t$
satisfies TC and IBC holds. Therefore, the Quintos (1995) concept of strong sustainability,
defined as $B_t\sim I(1)$,
while intuitively appealing, is one of many possibilities of data generating
processes satisfying IBC.

Second, there are economic considerations that
restrict the DGP, besides IBC. For example, fiscal sustainability may involve
bounds or restrictions on the deficit $\Delta B_t$ that can be formulated as $\Delta B_t\sim I(0)$,
which corresponds to strong sustainability by Quintos (1995),
and $G_t - R_t\sim I(0)$ if $G_t, R_t\sim I(1)$. Furthermore,
there could be bounds on deviations of debt from revenue, that can
be formulated as cointegration between $B_t$ and $R_t$. In that
case $G_t$ and $R_t$ are multicointegrated and the conditions for the asymptotic
result in Phillips and Hansen
(1990) employed in Quintos (1995) are not met. To allow for multicointegration, Berenguer-Rico and
Carrion-i-Silvestre (2011) model the revenue-expenditure relationship in an $I(2)$ VAR
system, as suggested by Haldrup (1994) and Engsted et al (1997). 
The results of the present paper show that it is not necessary to work in an $I(2)$ system and it is possible to go beyond a VAR specification. In particular: (i) multicointegration can be allowed directly in the
$I(1)$ system considered in Equation (6) in Quintos (1995); (ii) multicointegration invalidates the 
normal approximation of the test statistics $t^{+}$ used in Section 3.1.2 in Quintos (1995); and (iii) multicointegration does not alter the conclusion that the null hypothesis of cointegration between $G_t$ and $R_t$ with coefficient $(1,-1)$ is rejected. We explore these points and provide revised estimates and tests based on an  updated dataset. 

The data are provided  by the US Bureau of Economic Analysis
and retrieved
from FRED, Federal Reserve Bank of St.{}~Louis on November 17, 2019. 
We consider two series: $x_t = $
Government Current Expenditures (GEXPND), inclusive of interest payments, and $y_t = $ Government Current Receipts
(GRECPT).
Both series are in
billions of dollars,
seasonally adjusted annual rate, at quarterly frequency from 1947:Q1 to
2019:Q1, $T=291$ observations.

\begin{figure}[!ht]
\centering
\subfigure[Levels] {
\includegraphics[width=0.47\textwidth]{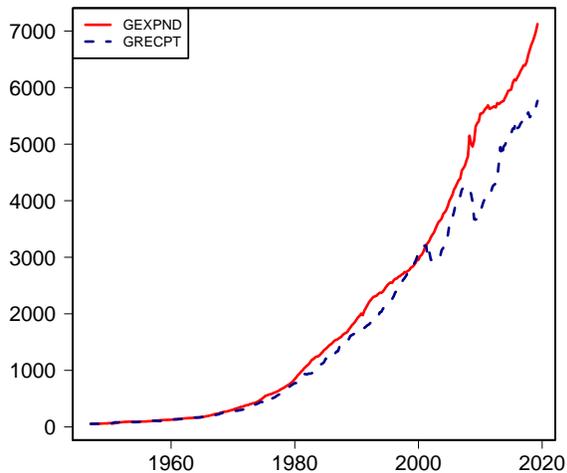}
}
\subfigure[Logs] {
\includegraphics[width=0.47\textwidth]{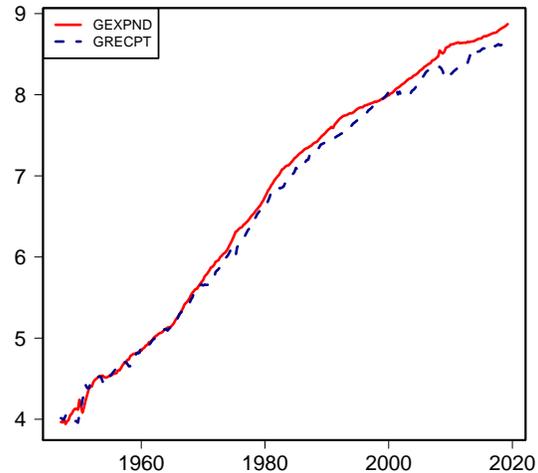}
}
\subfigure[First differences] {
\includegraphics[width=0.47\textwidth]{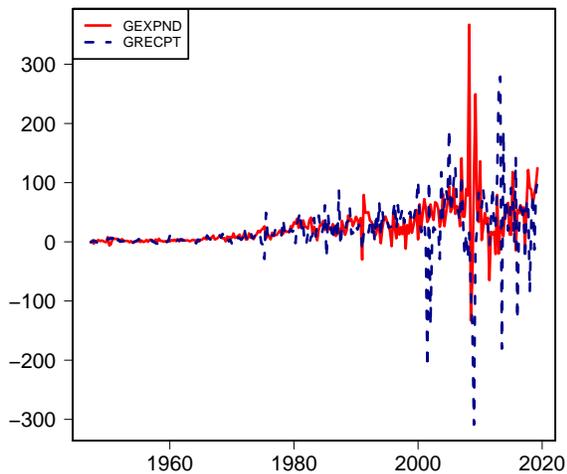}
}
\subfigure[First differences of logs] {
\includegraphics[width=0.47\textwidth]{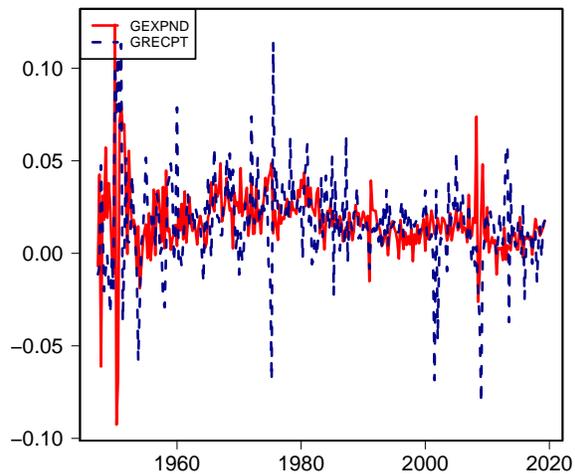}
} 
\caption{US Government expenditures and receipts, billions of dollars,
seasonally adjusted annual rate, quarterly frequency.}\label{fig:ER}
\end{figure}

The series are plotted in Figure~\ref{fig:ER}(a). We see that the series start to
diverge in the mid 1990s and even more so after year 2000.  We estimate the equation $y_t=A x_t
+u_{0t}$ and test the null hypothesis of strong sustainability, viz., $\mathcal{H}_0:\ A=1$.
FM-OLS estimation of the full sample gives $\widehat A^{+} = 0.83$ with
standard error $0.01$ and $t$-statistic $(0.83-1)/0.01=-17$, rejecting the null hypothesis.
The result is similar if we include the constant and for bandwidth $T^{1/5}$ in
place of $3T^{1/5}$.

The divergence of the series in mid 1990s in Figure~\ref{fig:ER}(a) may signify a
structural break in the relationship. In fact, several studies (e.g. Berenguer-Rico and
Carrion-i-Silvestre, 2011)  found a break in
the 4th quarter of 1996, which could be attributed to the 1997 Clinton tax cut.   The study of the properties of the  FM-OLS under
multicointegration in the presence of structural breaks we leave for future
research. But we do estimate the model for the period from 1947:Q1 to
1996:Q4 ($T=200$) finding that $\widehat A^{+} = 0.87$ with
standard error $0.005$ and $t$-statistic $(0.87-1)/0.005=-26$, so the
cointegrating coefficient is closer to but still statistically different from
$(1,-1)$.

From the Campbell--Shiller work on log-linearization of present value identities
we may expect that linear time series models provide better approximations in logarithms of the time series, which has the further advantage of stabilizing variances. 
The series in logs are plotted in Figure~\ref{fig:ER}(b). We also plot the first
differences in levels and in logs in Figure~\ref{fig:ER}. The first differences
of logs (Figure~\ref{fig:ER}(d)) show less heteroskedasticity than first
differences in levels (Figure~\ref{fig:ER}(c)) so our theory
results seem better suited for specification in logs\footnote{We thank a referee for
this suggestion.}. 
FM-OLS estimation of the full sample in logs gives $\widehat A^{+} = 0.98$ with
standard error $0.002$ and $t$-statistic $(0.98-1)/0.002=-10$, rejecting the
null hypothesis. The value of the $t$-statistic is similar for the period from
1947:Q1 to 1996:Q4.

We also estimate the cointegration relationship between real
revenue and expenditure constructed using the GDP deflator. We take the same
data series\footnote{Available at the
\textit{Journal of Applied Econometrics} Data Archive,
http://qed.econ.queensu.ca/jae/2011-v26.2/} as in Berenguer-Rico and Carrion-i-Silvestre (2011), but instead of looking at $I(2)$
systems  (which means working with $\sum_{j=0}^t R_j$, $\sum_{j=0}^t G_j$ and
$G_t$) we again run FM-OLS $R_t$ on $G_t$ and obtain $\widehat A^{+} = 0.92$ with
 standard error $0.01$ and $t$-statistic $(0.92-1)/0.01=-8$,
rejecting the null hypothesis that revenue and expenditure are cointegrated with
coefficient $(1,-1)$.

\section{Conclusion}
In a semiparametric triangular representation of $I(1)$ cointegrated time series
the presence of multicointegration results in a singular long run error variance matrix which has decisive effects on standard methods of estimation and inference in such models.  The consequences are higher rates of convergence and non pivotal limit theory in certain directions for estimators such as FM-OLS. Notwithstanding these effects, we show that FM-OLS Wald
tests are invariant to singularity under well defined rank conditions and, when those
conditions fail, the tests are conservative in certain cases. In particular, simulation experiments show that in such situations the test rejection rates under the null hypothesis are far below nominal levels based on standard asymptotics in singular and near singular cases. 
We illustrate our methods by analyzing the fiscal sustainability of the US
government, testing the hypothesis that government revenue and expenditure
are strongly cointegrated with coefficient $(1,-1)$, where multicointegration naturally
arises if bounds are imposed on deviations of debt from revenue. 

The results obtained here motivate the development of new robust approaches to
estimating cointegrating relationships that allow for the possible presence of multicointegration and that are pivotal in the presence of the singularity it produces. This is an ongoing area of research by the authors.

\section{Acknowledgments}
This paper has origins in a 2011 Yale Take Home Examination. For
various and non-overlapping parts of this research Kheifets
acknowledges support from the Russian Science Foundation under project
20-78-10113 (Monte Carlo simulations and the fiscal sustainability
evaluation) and the Spanish Ministerio de Ciencia, Innovacion y
Universidades under Grant ECO2017-86009-P (econometric theory).
Phillips acknowledges support from the National Science Foundation under Grant SES 18-50860
and a Kelly Fellowship at the University of Auckland. This research
was supported in part through computational resources of HPC
facilities at HSE University.

%\newpage

\appendix
\section{Appendix}

\subsection{Preliminary Lemmas}
We start by stating and proving some results for the rates of convergence and
limits of the  kernel estimator of the
long run variance when $\Omega_{00.x}=0$. These results are of independent
interest and are formulated as separate lemmas. The discussion proceeds and results are stated under the assumptions made in the
body of the paper.

The results contribute to general literature on the asymptotic bias and variance of spectral
estimates, see e.g.~Section~V in Hannan (1970).
We use ideas from  
Lemma 8.1 (a), (b), and (g) in Phillips (1995), 
although that lemma does not directly apply to our case.
In particular, the $I(-1)$ errors that appear in Lemma 8.1 in Phillips (1995) arise from a different source: if the regressor vector $x_t$ is cointegrated, but the cointegrating
relationship is unknown, FM-OLS uses the first differences of the full vector
$x_t$ in making nonparametric adjustments to OLS, thereby producing linear combination of the first differences of stationary errors which are $I(-1)$. In our case it is assumed that $\Omega_{xx}$ is
positive definite, i.e.~that  $x_t$ are full rank nonstationary $I(1)$ and
$\Delta{x_t}$ are full rank stationary $I(0)$. Instead, multicointegration induces singularity in the augmented regression equation error $u_{0.x,t}=u_{0t}-\Omega_{0x}\Omega_{xx}^{-1}u_{xt}$ in~\eqref{ppA} so that $u_{0.x,t}$ is $I(-1)$ with consequential effects on the estimation of the long run covariance matrix $\Omega_{00.x}$.  

Consider the case where $\Omega_{00.x}=0$. Define
\begin{align*}
\check\Delta_{\Delta e,u}=\sum_{j=0}^{T-1}w(j/K)\widehat\Gamma_{\Delta e,u}(j),
\quad
\Delta_{\Delta e,u}=\sum_{j=0}^{\infty}\Gamma_{\Delta e,u}(j),
\\
\check\Omega_{\Delta e,u}=\sum_{j=-T+1}^{T-1}w(j/K)\widehat\Gamma_{\Delta e,u}(j),
\quad
\Omega_{\Delta e,u}=\sum_{j=-\infty}^{\infty}\Gamma_{\Delta e,u}(j),\\
\quad\text{and}\quad
\check\Omega_{\Delta e,\Delta e}=\sum_{j=-T+1}^{T-1}w(j/K)\widehat\Gamma_{\Delta e,\Delta e}(j),
\end{align*}
where $w(\cdot)$ is a kernel function, $K$ is a bandwidth parameter,
$\Gamma_{a,b}(j)=\E a_{t+j} b_{t}^{\prime}$,
$\widehat\Gamma_{a,b}(j)
=T^{-1}\sum_{1\le t,t+j\le T} a_{t+j} b_{t}^{\prime}$,
and we use  $\ \check{}\ $  in place of $\ \widehat{}\ $ to stress that the
kernel estimators are calculated with the true errors $e_t$ and $u_t$.
\begin{lemma} \label{lem:eu}
Suppose $\Omega_{00.x}=0$. Then
\begin{enumerate}[(a)]
\item 
$
\check\Delta_{\Delta e,u}+\widehat\Gamma_{e,u}(-1)
=
-K^{-2}w''(0)\sum_{j=0}^{\infty}\left(j+1/2\right)\Gamma_{e,u}(j)\\
\quad{}+O_p\left(\left(KT\right)^{-1/2}\right)
+o_p\left(K^{-2}\right),
$ and \\
$
\check\Delta_{\Delta e,u}-\Delta_{\Delta e,u}
=
-K^{-2}w''(0)\sum_{j=0}^{\infty}\left(j+1/2\right)\Gamma_{e,u}(j)\\
\quad{}+O_p\left(T^{-1/2}\right)
+o_p\left(K^{-2}\right),
$
\item %(b)
$
\check\Omega_{\Delta e,u}
=
-K^{-2}w''(0)\sum_{j=-\infty}^{\infty}\left(j+1/2\right)\Gamma_{e,u}(j)
+O_p\left(\left(KT\right)^{-1/2}\right)
+o_p\left(K^{-2}\right),
$
\item %(c)
$
\check\Omega_{\Delta e,\Delta e}
=
-K^{-2}w''(0)\sum_{j=-\infty}^{\infty}\Gamma_{e,e}(j)
+o_p\left(K^{-2}\right).
$
\end{enumerate}
\end{lemma}

Next consider the case when the kernel
estimators are based on regression residuals $(\widehat u'_{0t},u'_{xt})$, where
$\widehat u'_{0t} = y_t -\widehat A x_t= u_{0t} -(\widehat A -A)x_t$, and using the true
transform $\Omega_{0x}\Omega_{xx}^{-1}$ define $\widetilde u_{0.x,t}=\widehat
u_{0t}-\Omega_{0x}\Omega_{xx}^{-1}u_{xt}$, whereas $\widehat u_{0.x,t}=\widehat
u_{0t}-\widehat\Omega_{0x}\widehat\Omega_{xx}^{-1}u_{xt}$. 
Define
$\Delta_{0x}^+=\sum_{j=0}^{\infty}\Gamma_{u_{0.x},u_x}(j)$,
\begin{align*}
&\widetilde\Delta_{0x}^{+}=\sum_{j=0}^{T-1}w(j/K)\widehat\Gamma_{\widetilde u_{0.x},u_x}(j),
&\check\Delta_{0x}^{+}=\sum_{j=0}^{T-1}w(j/K)\widehat\Gamma_{u_{0.x},u_x}(j),\\
&\widetilde\Omega_{0x}^{+}=\sum_{j=-T+1}^{T-1}w(j/K)\widehat\Gamma_{\widetilde
u_{0.x},u_x}(j),
&\check\Omega_{0x}^{+}=\sum_{j=-T+1}^{T-1}w(j/K)\widehat\Gamma_{
u_{0.x},u_x}(j),\\
&\widetilde\Omega_{00.x}=\sum_{j=-T+1}^{T-1}w(j/K)\widehat\Gamma_{\widetilde
u_{0.x},\widetilde u_{0.x}}(j),
&\check\Omega_{00.x}=\sum_{j=-T+1}^{T-1}w(j/K)\widehat\Gamma_{
u_{0.x},u_{0.x}}(j),
\end{align*}
where we use  $\ \widetilde{}\ $  in place of $\ \check{}\ $ to stress that the
kernel estimators use residuals $(\widetilde u_{0.x,t}',u_{xt}')'$ instead of the true
errors $(u_{0.x,t}',u_{xt}')'$. 

The following lemmas hold irrespective of the singularity of $\Omega_{00.x}.$
\begin{lemma} \label{lem:0x}
$\widetilde\Delta_{0x}^{+}-\check\Delta_{0x}^{+}$,
$\widetilde\Omega_{0x}^{+}-\check\Omega_{0x}^{+}$,
$\widetilde\Omega_{00.x}-\check\Omega_{00.x}$ are $O_p(K/T)$.
\end{lemma}

\begin{lemma}\label{lem:UX}
If $\Omega_{00.x}=0$, then for $u_{0.x,t}=-\Delta e_t$,
\begin{enumerate}[(a)]
\item $
T^{-1}U'_{0.x} X-\widetilde\Delta_{0x}^{+}
=T^{-1}e_T x'_T +
K^{-2}w''(0)\sum_{j=0}^{\infty}\left(j+1/2\right)\Gamma_{e,u_x}(j)\nonumber\\
\quad{}+O_p\left(\left(KT\right)^{-1/2}\right)+O_p(K/T)
+o_p\left(K^{-2}\right),
$
\item $
\widetilde\Omega_{0x}^{+}
=K^{-2}w''(0)\sum_{j=-\infty}^{\infty}\left(j
+1/2\right)\Gamma_{e,u_x}(j)\nonumber\\
\quad{}+O_p\left(\left(KT\right)^{-1/2}\right)+O_p(K/T)
+o_p\left(K^{-2}\right),
$
\item $
\widetilde\Omega_{00.x}
=-K^{-2}w''(0)\sum_{j=-\infty}^{\infty}\Gamma_{e,e}(j)
+O_p(K/T)
+o_p\left(K^{-2}\right).
$
\end{enumerate}
\end{lemma} 

\subsection{Proofs of Lemmas 1 - 3}
\begin{proof}[Proof of Lemma \ref{lem:eu}(a)]
By Assumption K, $w(0)=1$ and $w(x)=0$ for $x\ge 0$, and 
for some $K^*=K^b$ with $0<b<1$
\begin{align*}
\check\Delta_{\Delta e,u}=&\sum_{j=0}^{T-1}w(j/K)\widehat\Gamma_{\Delta e,u}(j),\\
=&\sum_{j=0}^{K-1}w(j/K)
\left(
\widehat\Gamma_{e,u}(j)
-\widehat\Gamma_{e,u}(j-1)
\right)\\
=&\left(\sum_{j=0}^{K^*}+\sum_{j=K^*+1}^{K-2}\right)
\left(w(j/K)-w((j+1)/K)\right)
\widehat\Gamma_{e,u}(j)
-\widehat\Gamma_{e,u}(-1)\\
&{}+w((K-1)/K)\widehat\Gamma_{e,u}(K-1)=\sum_{k=1}^4 R_k,\ \text{say}.
\end{align*}
We show that 
$R_1+R_2=-K^{-2}w''(0)\sum_{j=0}^{\infty}\left(j+1/2\right)\Gamma_{e,u}(j)
+O_p\left(\left(KT\right)^{-1/2}\right)
+o_p\left(K^{-2}\right)$,
$R_3=-\widehat\Gamma_{e,u}(-1)=\Delta_{\Delta e,u} + O_p\left(T^{-1/2}\right)$, and 
$R_4=O_p\left(K^{-2}T^{-1/2}\right)$.

\noindent\textbf{Mean of $\mathbf{R_1}$.}
Applying the second order Taylor expansion of function $w(\cdot)$ at arguments
$(j+1)/K$
around $j/K$,
\begin{align*}
w((j+1)/K)-w((j)/K)=K^{-1}w'(j/K)+1/2\ K^{-2}w''(j/K)[1+o(1)],
\end{align*}
and for $j\le K^*$ we can apply the Taylor expansion of function $w'(\cdot)$ at
arguments $j/K$ around $0$, where $w'(0)=0$, giving $w'(j/K)=w''(0)(j/K)[1+o(1)],$
and then
$w((j+1)/K)-w(j/K)=K^{-2}w''(0)(j+1/2)[1+o(1)].$
Hence, 
\begin{align*}
K^2  R_1 =& K^2\sum_{j=0}^{K^*}[w(j/K)-w((j+1)/K)]
\widehat\Gamma_{e,u}(j) 
= -w''(0)
\sum_{j=0}^{K^*}(j+1/2)
\widehat\Gamma_{e,u}(j) [1+o(1)],
\end{align*}
and
\begin{align*}
\E
\sum_{j=0}^{K^*}(j+1/2)
\widehat\Gamma_{e,u}(j)
=& \sum_{j=0}^{K^*}(j+1/2)
\left(1-j/T\right)\Gamma_{e,u}(j) 
\to
\sum_{j=0}^{\infty}(j+1/2)
\Gamma_{e,u}(j).
\end{align*}

\noindent\textbf{Mean of $\mathbf{R_2}.$} 
By the mean value theorem there exists $x_{j,K}\in(j/K,(j+1)/K)$, such that
$R_2 =K^{-1}
\sum_{j=K^* +1}^{K-2}
w'(x_{j,K})\widehat\Gamma_{e,u}(j)$
with mean
\begin{align*}
\E R_2=K^{-1}
\sum_{j=K^* +1}^{K-2}
w'(x_{j,K})(1-j/T)\Gamma_{e, u}(j).
\end{align*}
whose modulus is dominated by
\begin{align*}
\sup_x & |w'(x)| K^{-1}
\sum_{j=K^* +1}^{K-2}\|\Gamma_{e,u}(j)\|\\
&\le const\ K^{-1}
\sum_{j>K^*}\sum_{s=0}^{\infty}
\|D_{s}\|\|\widetilde D_{s+j}\|\\
&\le const\ K^{-1}K^{*^{-\nu}}
\sum_{j>K^*}\sum_{s=0}^{\infty}
(s+j)^{\nu}\|D_{s}\|\|D_{s+j}\|\\
&\le const\ K^{-1}K^{{-\nu b}}
\sum_{s=0}^{\infty}\|D_{s}\|
\sum_{r=0}^{\infty}r^{\nu}\|D_{r}\|
=O(K^{-1-\nu b})=o(K^{-2}),
%\label{eq:pKstars}
\end{align*}
for $1/\nu<b<1$.

\noindent\textbf{Variance of $\mathbf{R_1+R_2}$.}
\begin{align*}
R_1+R_2=
\sum_{j=0}^{K-2}
\left(w\left(\frac{j}{K}\right)-w\left(\frac{j+1}{K}\right)\right)
\widehat\Gamma_{e,u}(j)
= K^{-1}\sum_{j=0}^{K-2}
w'\left(\frac{j}{K}\right)\widehat\Gamma_{e,u}(j)[1+O(K^{-1})]
\end{align*}
and $\mathbb{V}ar\left[vec\left[
\sum_{j=0}^{K-2}w'(j/K)\widehat\Gamma_{e,u}(j)
\right]\right]=O(KT^{-1})$
from Theorem~9 in Hannan (1970, p.~280).
So the variance of the dominant term in $R_1+R_2$ is $O(K^{-1}T^{-1})$.

\noindent\textbf{Term $\mathbf{R_3}$.}
Note that
$\Delta_{\Delta e,u}=\sum_{j=0}^{\infty}\Gamma_{\Delta
e,u}(j)=-\Gamma_{e,u}(-1)$.
$\E R_3=-\E \widehat\Gamma_{e,u}(-1)=-(1-1/T)\Gamma_{e,u}(-1)=(1-1/T)\Delta_{\Delta
e,u}$, 
and $\mathbb{V}ar(\widehat\Gamma_{e,u}(-1))=O(T^{-1})$.
Therefore, $-\widehat\Gamma_{e,u}(-1)=\Delta_{\Delta e,u}+O(T^{-1/2})$.

\noindent\textbf{Term $\mathbf{R_4}$.}
By Assumption K (b), $w((K-1)/K)=O(K^{-2})$ when $K\to\infty$. By
summability, 
$\E \widehat\Gamma_{e,u}(K-1)=(1-(K-1)/T)\Gamma_{e,u}(K-1)=o(1)$
and $\mathbb{V}ar(\widehat\Gamma_{e,u}(K-1))=O(T^{-1})$.
Therefore, $\widehat\Gamma_{e,u}(K-1)=O_p(T^{-1/2})$
and $R_4=O_p(K^{-2}T^{-1/2})$.
\end{proof}

\begin{proof}[Proof of Lemma \ref{lem:eu}(b)]
As above,
\begin{align*}
&\check\Omega_{\Delta e,u}
=\sum_{j=-T+1}^{T-1}w(j/K)\widehat\Gamma_{\Delta e,u}(j),
=\sum_{j=-K+1}^{K-1}w(j/K)
\left(
\widehat\Gamma_{e,u}(j)
-\widehat\Gamma_{e,u}(j-1)
\right)\\
=&\left(\sum_{j=-K^*}^{K^*}+\sum_{|j|=K^*+1}^{K-2}\right)
\left(w(j/K)-w((j+1)/K)\right)
\widehat\Gamma_{e,u}(j)\\
&{}-w((-K+1)/K)\widehat\Gamma_{\Delta e,u}(-K)
+w((K-1)/K)\widehat\Gamma_{\Delta e,u}(K-1)\\
&{}=\sum_{k=1}^4 R_k,\ \text{say}.
\end{align*}
and
$R_1+R_2=K^{-2}w''(0)\sum_{j=-\infty}^{\infty}\left(j+1/2\right)\Gamma_{e,u}(j)+O_p\left(\left(KT\right)^{-1/2}\right)
+o_p\left(K^{-2}\right)$ while  $R_k=O_p(K^{-2}T^{-1/2})$, for $k=3,4$ by
the same argument as in (a).
\end{proof}

\begin{proof}[Proof of Lemma \ref{lem:eu}(c)]
Similar to (b), for some $K^*=K^b$ with $0<b<1$
\begin{align*}
&\check\Omega_{\Delta e,\Delta e}
=\sum_{j=-T+1}^{T-1}w(j/K)\widehat\Gamma_{\Delta e,\Delta e}(j),
=\sum_{j=-K+1}^{K-1}w(j/K)
\left(
\widehat\Gamma_{e,\Delta e}(j)
-\widehat\Gamma_{e,\Delta e}(j-1)
\right)\\
=&\left(\sum_{j=-K^*}^{K^*}+\sum_{|j|=K^*+1}^{K-2}\right)
\left(w(j/K)-w((j+1)/K)\right)
\widehat\Gamma_{e,\Delta e}(j)\\
&{}-w((-K+1)/K)\widehat\Gamma_{\Delta e,\Delta e}(-K)
+w((K-1)/K)\widehat\Gamma_{\Delta e,\Delta e}(K-1)\\
&{}=\sum_{k=1}^4 R_k,\ \text{say}.
\end{align*}
By the same argument as above, $\E R_2=o\left(K^{-2}\right)$ and
$R_k=O_p(K^{-2}T^{-1/2})$, for $k=3,4$, 
while we show below that the limit of $R_1$ can be simplified and the variance bound of $R_1+R_2$
can be improved from $O(K^{-1}T^{-1})$ in (a) and (b) to
$O(K^{-3}T^{-1})$ in the present case. Thus, the contribution from the variance
is of smaller order than $K^{-2}$ for any $K=T^k$, with $k\in(0,1)$.

\noindent\textbf{Mean of $\mathbf{R_1}$.}
\begin{align*}
K^2  R_1 =& K^2\sum_{|j|<K^*}[w(j/K)-w((j+1)/K)]
\widehat\Gamma_{e,\Delta e}(j) \\
&=
-w''(0)
\sum_{|j|<K^*}(j+1/2)
\widehat\Gamma_{e,\Delta e}(j) [1+o(1)],
\end{align*}
and
\begin{align*}
\E\sum_{|j|<K^*}(j+1/2)\widehat\Gamma_{e,\Delta e}(j)
=& \sum_{|j|<K^*}(j+1/2)
\left(1-j/T\right)\Gamma_{e,\Delta e}(j) \\
&\to
\sum_{j=-\infty}^{\infty}(j+1/2)\Gamma_{e,\Delta e}(j)
=\sum_{j=-\infty}^{\infty}\Gamma_{e, e}(j).
\end{align*}
because
$\sum_{j=-\infty}^{\infty}j\Gamma_{e,\Delta e}(j)
=\sum_{j=-\infty}^{\infty}j\Gamma_{e, e}(j)
-\sum_{j=-\infty}^{\infty}j\Gamma_{e, e}(j+1)
=\sum_{j=-\infty}^{\infty}\Gamma_{e,e}(j)$
and
$\sum_{j=-\infty}^{\infty}\Gamma_{e,\Delta e}(j)=0$.
Note that under stricter conditions on the bandwidth rate this limit could be obtained from Theorem 10 in Hannan (1970, p.
283).

\noindent\textbf{Variance of $\mathbf{R_1+R_2}$.}
\begin{align*}
R_1+R_2=&
\sum_{j=-K+1}^{K-2}
\left(w(j/K)-w((j+1)/K)\right)
\widehat\Gamma_{e,\Delta e}(j)\\
&=
-K^{-1}\sum_{j=-K+1}^{K-2}
w'(j/K)\widehat\Gamma_{e,\Delta e}(j)[1+O(K^{-1})]\\
&=
-K^{-2}\sum_{j=-K+2}^{K-2}
w''((j-1)/K)\widehat\Gamma_{e, e}(j)+o_p(K^{-2})
\end{align*}
and $\mathbb{V}ar\left[vec\left[
\sum_{j=-K+2}^{K-2}w''((j-1)/K)\widehat\Gamma_{e, e}(j)
\right]\right]=O(KT^{-1})$
from Theorem~9 in Hannan (1970, p.280).
The variance of the dominant term in $R_1+R_2$ is therefore
$O(K^{-4}KT^{-1})=O(K^{-3}T^{-1})$.
\end{proof}

\begin{proof}[Proof of Lemma \ref{lem:0x}]
Because
$\widetilde u_{0.x,t}-u_{0.x,t}=\widehat u_{0t} - u_{0t} = -(\widehat A-A)x_t$
\begin{align*}
&\widetilde\Delta_{0x}^{+}-\check\Delta_{0x}^{+}
=-(\widehat A-A)\sum_{j=0}^{T-1}w(j/K)\widehat\Gamma_{x,u_x}(j)=O_p(K/T),\\
&\widetilde\Omega_{0x}^{+}-\check\Omega_{0x}^{+}
=-(\widehat A-A)\sum_{j=-T+1}^{T-1}w(j/K)\widehat\Gamma_{x,u_x}(j)=O_p(K/T),\\
&\widetilde\Omega_{00.x}-\check\Omega_{00.x}
=(\widehat A-A)\sum_{j=-T+1}^{T-1}w(j/K)\widehat\Gamma_{x,x}(j)(\widehat A-A)'\\
&-(\widehat A-A)\sum_{j=-T+1}^{T-1}w(j/K)\widehat\Gamma_{x,u_{0.x}}(j)
-\sum_{j=-T+1}^{T-1}w(j/K)\widehat\Gamma_{u_{0.x}, x}(j)(\widehat
A-A)'\\
{}&=O_p(K/T),
\end{align*}
because $\widehat A-A=O_p(T^{-1})$ and 
it follows from the proof of Theorem 3.1 of Phillips (1991b, pp.~432--433) that
\begin{align*}
&\sum_{j=0}^{T-1}w(j/K)\widehat\Gamma_{x,u_x}(j)=O_p(K),
\sum_{j=-T+1}^{T-1}w(j/K)\widehat\Gamma_{x,u_x}(j)=O_p(K),\\
&\sum_{j=-T+1}^{T-1}w(j/K)\widehat\Gamma_{x,x}(j)=O_p(KT),
\sum_{j=-T+1}^{T-1}w(j/K)\widehat\Gamma_{u_{0.x},x}(j)=O_p(K).
\end{align*}
If $\Omega_{00.x}=0$, the last bound can be improved to $O_p(1)$
\begin{align*}
&\sum_{j=-T+1}^{T-1}w(j/K)\widehat\Gamma_{u_{0.x},x}(j)
=
\sum_{j=-T+1}^{T-1}w(j/K)\widehat\Gamma_{\Delta e,x}(j)
=O_p(1)
\end{align*}
similar to the proof of Lemma 8.1(c) of Phillips (1995, pp.~1064--1065).
\end{proof}

\begin{proof}[Proof of Lemma \ref{lem:UX}]
We apply Lemma~\ref{lem:eu} to  the covariance between $U_{0.x}$
and $U_x$ and
 adjust the limits for the use of residual $\widehat U_{0}$ in place of the true error
$U_0$ with an additional term of $O_p(K/T)$ according to Lemma~\ref{lem:0x}, and obtain parts
(b) and~(c). Part~(a) follows
by partial summation formula and applying the above procedure to the term in the
brackets:
\begin{align*}
T^{-1}U'_{0.x} X-\widetilde\Delta_{0x}^{+}
&=T^{-1}e_T x'_T -
\left(\widetilde\Delta_{0x}^{+}+\widehat\Gamma_{e,u_x}(-1)\right)\\
&=T^{-1}e_T x'_T+
K^{-2}w''(0)\sum_{j=0}^{\infty}\left(j+1/2\right)\Gamma_{e,u_x}(j)\nonumber\\
&\quad{}+O_p\left(\left(KT\right)^{-1/2}\right)+O_p(K/T)
+o_p\left(K^{-2}\right).
\end{align*}
\end{proof}
\subsection{Proofs of Propositions}
\begin{proof}[Proof of Proposition \ref{prop:mcoint}]
We can write $\left(y'_t,x'_t\right)'=\left(1-L\right)^{-1}C(L)\eta_t$, where the roots of
$|C(z)|=0$ satisfy $|z|> 1$ or $z=1$.
Multicointegration of such a linear
$I(1)$ process occurs (see Johansen 1992, Engsted and Johansen, 1999)
when $z=1$ is a root, so that $C(1)=\xi\epsilon'$ has reduced rank and $\xi'_{\perp}\dot
C(1)\epsilon_{\perp}$ is singular. The submatrices $\{\xi,\epsilon\}$ 
and their orthogonal complements $\{\xi_{\perp},\epsilon_{\perp}\}$are given explicitly below for the present context of the triangular system \eqref{pp0} and \eqref{pp1}. In particular, write $C(L)$ as
\begin{equation*}
C(L)=\left[
\begin{array}{cc}
(1-L)I_{m_0} & A \\
0 & I_{m_x}%
\end{array}%
\right]D(L),
\end{equation*}
and its derivative
\begin{equation}\label{eq:dotC}
\dot C(L)=\left[
\begin{array}{cc}
-I_{m_0} & 0 \\
0 & 0%
\end{array}%
\right]D(L)
+
\left[
\begin{array}{cc}
(1-L)I_{m_0} & A \\
0 & I_{m_x}%
\end{array}%
\right]\dot D(L).
\end{equation}
We can write
\begin{equation*}
C(1)=\left[
\begin{array}{cc}
0 & A \\
0 & I_{m_x}%
\end{array}%
\right]D(1)
=
\left[
\begin{array}{cc}
A  \\
I_{m_x}%
\end{array}%
\right]
[D_{x0}(1),D_{xx}(1)]
=\xi\epsilon',\quad\text{taking}\quad
\xi=
\left[
\begin{array}{cc}
A  \\
I_{m_x}%
\end{array}%
\right],
\end{equation*}
$\xi'_{\perp}=[I_{m_x},-A]$ and $\epsilon'= [D_{x0}(1), D_{xx}(1)]$, which is an $m_x\times m$ matrix of full rank $m_x$ whenever $\epsilon'\epsilon=\Omega_{xx}>0$.
Multiplication of $D(1)$ by the full rank $m\times m$ matrix
$\epsilon^F=[\epsilon_{\perp},\epsilon]$ yields
\begin{align*}
rank(D(1))=rank(D(1)\epsilon^F)
=rank(\xi'_{\perp}\dot C(1)\epsilon_{\perp}) + rank(\Omega_{xx}),
\end{align*}
because
\begin{align*}
D(1)\epsilon^F=
\left[
\begin{array}{cc}
[D_{00}(1), D_{0x}(1)]\epsilon_{\perp} & [D_{00}(1),
D_{0x}(1)]\epsilon\\
0 & \Omega_{xx}%
\end{array}%
\right],
\end{align*}
and from (\ref{eq:dotC})
\begin{equation*}
\xi'_{\perp}\dot C(1)\epsilon_{\perp}
=
\left[
\begin{array}{cc}
-I_{m_0} & 0
\end{array}%
\right]D(1)\epsilon_{\perp}
=
-[D_{00}(1), D_{0x}(1)]\epsilon_{\perp},
\end{equation*}
since
\begin{equation*}
\xi'_{\perp}
\left[
\begin{array}{cc}
-I_{m_0} & 0 \\
0 & 0%
\end{array}%
\right]
=
\left[
\begin{array}{cc}
-I_{m_0} & 0
\end{array}%
\right]
\quad\text{and}\quad
\xi'_{\perp}
\left[
\begin{array}{cc}
0 & A \\
0 & I_{m_x}%
\end{array}%
\right]
=0.
\end{equation*}
It follows that when $rank(\Omega_{xx})=m_x$, we have the rank equivalence $m_0 - rank(\xi'_{\perp}\dot C(1)\epsilon_{\perp})=m-rank(\Omega)=m_0-rank(\Omega_{00.x})$, which is the multicointegrating rank in the system defined by \eqref{pp0} and \eqref{pp1}. More simply, $rank(\xi'_{\perp}\dot C(1)\epsilon_{\perp})=rank(\Omega_{00.x})$. 
\end{proof}

\begin{proof}[Proof of Proposition \ref{prop:Rfull}]
Recall that
$\widetilde\Omega_{0x}^{+}=\widehat\Omega_{0x} -
\Omega_{0x}\Omega_{xx}^{-1}\widehat\Omega_{xx}$, so
\begin{align*}
\widehat\Omega_{0x}\widehat\Omega_{xx}^{-1} - \Omega_{0x}\Omega_{xx}^{-1}
=\left(\widehat\Omega_{0x} - \Omega_{0x}\Omega_{xx}^{-1}\widehat\Omega_{xx}\right)\widehat \Omega^{-1}_{xx}
=\widetilde\Omega_{0x}^{+}\widehat \Omega^{-1}_{xx}.
\end{align*} 
Also,
$\widetilde\Delta_{0x}^{+}=\widehat\Delta_{0x} -
\Omega_{0x}\Omega_{xx}^{-1}\widehat\Delta_{xx}$, so
\begin{align*}
\widehat\Delta^{+}_{0x}
=\widetilde\Delta_{0x}^{+}-\left(\widehat\Omega_{0x}\widehat\Omega_{xx}^{-1} - \Omega_{0x}\Omega_{xx}^{-1}\right)
\widehat\Delta_{xx}
=
\widetilde\Delta_{0x}^{+}-\widetilde\Omega_{0x}^{+}\widehat\Omega^{-1}_{xx}
\widehat\Delta_{xx}.
\end{align*} 
The numerator matrix of FM-OLS is
\begin{align*}
&\widehat Y^{+}{}'X-T\widehat\Delta^{+}_{0x}
=\left(Y'- \widehat\Omega_{0x}\widehat\Omega_{xx}^{-1} U_{x}'\right)X
-T\widehat\Delta^{+}_{0x}\\
{ }&=AX'X + U'_{0}X - \widehat\Omega_{0x}\widehat\Omega_{xx}^{-1}
U_{x}'X-T\widehat\Delta^{+}_{0x}\\
{}&=AX'X + U'_{0.x}X
-T\widetilde\Delta^{+}_{0x}-\left(\widehat\Omega_{0x}\widehat\Omega_{xx}^{-1} -
\Omega_{0x}\Omega_{xx}^{-1}\right) \left(U_{x}'X-T\widehat\Delta_{xx}\right)\\
{}&=AX'X + 
U'_{0.x}X
-T\widetilde\Delta^{+}_{0x}-
\widetilde\Omega_{0x}^{+}\widehat\Omega^{-1}_{xx}
\left(U_{x}'X-T\widehat\Delta_{xx}\right).
\end{align*} 
Therefore,
\begin{align}%\label{eq:expandA}
T\left(\widehat A^{+} - A\right)
=&\left(T^{-1}U'_{0.x} X -
\widetilde\Delta_{0x}^{+}\right)\left(T^{-2}X'X\right)^{-1}\nonumber\\
&\quad{}-\widetilde\Omega_{0x}^{+}\widehat \Omega^{-1}_{xx}\left(T^{-1}U'_x
X-\widehat\Delta_{xx}\right)\left(T^{-2}X'X\right)^{-1}\label{eq:TAplus-A}.
\end{align} 
From standard weak convergence theory for sample covariances in
Phillips and Durlauf (1986) and Phillips (1989)
\begin{align}
T^{-2}X'X &\to_d \int_0^1 B_x B'_x,\label{eq:XX}\\
T^{-1}U'_x X &\to_d \int_0^1 d B_x B'_x + \Delta_{xx},\label{eq:UxX}\\
T^{-1}U'_0 X &\to_d \int_0^1 d B_0 B'_x +\Delta_{0x}\label{eq:U0X},
\end{align}
giving 
\begin{align*}\label{eq:wrate}
T^{-1}U'_{0.x} X\to_d \int_0^1 d B_{0.x} B'_x+ \Delta_{0x}^{+}.
\end{align*} 
By construction, $u_{0.x,t}$ has zero long run covariance with the errors $u_{xt}$
that drive the nonstationary component $x_t$, thereby eliminating the endogeneity from $x_t$ in the
long run. Therefore, with any consistent estimators of
$\Omega$ and $\Delta$, under Assumption K with $0<k<1$, and using 
Lemma~2, we have $\widetilde\Delta_{0x}^{+}\to_p\Delta_{0x}^{+}$
and
$\widetilde\Omega_{0x}^{+}\to_p 0$,
so that by continuous mapping and joint convergence of the components we have
\begin{align*}
T\left(\widehat A^{+} - A\right)\to_d
\left(\int_0^1 d B_{0.x} B'_x\right) \left(\int_0^1 B_x B'_x\right)^{-1}.
\end{align*} 
\end{proof}
\begin{proof}[Proof of Proposition \ref{prop:Rnull}]
The proof follows from expansion~(\ref{eq:TAplus-A}) by using the consistency of the
kernel estimates, the convergences~(\ref{eq:XX}) and (\ref{eq:UxX}), and the rates of
convergence  established in Lemma~\ref{lem:UX}(a) and (b).
\end{proof}
\begin{proof}[Proof of Proposition \ref{prop:Rnulldistr}]
For $k<1/4$, $O_p(K^{-2})$ dominates
$O_p(K^{-2})+O_p\left((KT)^{-1/2}\right)+O_p(K/T)$ and the limits in
Lemma~\ref{lem:UX}(a) and (b) have the form
\begin{align*}
T^{-1}U'_{0.x} X-\widetilde\Delta_{0x}^{+}
&=
K^{-2}w''(0)\sum_{j=0}^{\infty}\left(j+1/2\right)\Gamma_{e,u_x}(j)
+o_p\left(K^{-2}\right),\\
\widetilde\Omega_{0x}^{+}
&=K^{-2}w''(0)\sum_{j=-\infty}^{\infty}\left(j+1/2\right)\Gamma_{e,u_x}(j)
+o_p\left(K^{-2}\right),
\end{align*}
which together with (\ref{eq:XX}) and (\ref{eq:UxX}) and
expansion~(\ref{eq:TAplus-A})
give the limit distribution of
$K^2 T\left(\widehat A^{+} - A\right)$. 
\end{proof} 
\begin{proof}[Proof of Proposition \ref{prop:RnullDenomDistr}]

\emph{Part~(a)}.
The rate of convergence and limit behavior of $\widetilde\Omega_{00.x}$ is
established in Lemma~\ref{lem:UX}(c), while the difference
$\widehat\Omega_{00.x}-\widetilde\Omega_{00.x}$ is of smaller order and can be neglected, viz., 
\begin{align*}
&\widehat\Omega_{00.x}
-\widetilde\Omega_{00.x}
=
-
\left(
\widehat\Omega_{0x}\widehat\Omega_{xx}^{-1}
-\Omega_{0x}\Omega_{xx}^{-1}
\right)
\widehat\Omega_{x0}
-
\left[
\left(
\widehat\Omega_{0x}\widehat\Omega_{xx}^{-1}
-\Omega_{0x}\Omega_{xx}^{-1}
\right)
\widehat\Omega_{x0}
\right]'\\
&+
\widehat\Omega_{0x}
\left(
\widehat\Omega_{0x}\widehat\Omega_{xx}^{-1}
-\Omega_{0x}\Omega_{xx}^{-1}
\right)'
+
\left(\widehat\Omega_{0x}\widehat\Omega_{xx}^{-1}
-\Omega_{0x}\Omega_{xx}^{-1}\right)
\widehat\Omega_{xx}^{-1}\Omega_{xx}^{-1}\Omega_{0x}'\\
&=
\left(\widehat\Omega_{0x}\widehat\Omega_{xx}^{-1}
-\Omega_{0x}\Omega_{xx}^{-1}\right)
\left(-\widehat\Omega_{0x}
+\widehat\Omega_{xx}^{-1}\Omega_{xx}^{-1}\Omega_{0x}'\right)
=-\widetilde\Omega_{0x}^{+}\widehat\Omega_{xx}^{-1}
\widetilde\Omega_{0x}^{+}{}'.
\end{align*}
By comparing the rates in Part~(b) and Part~(c) in Lemma~\ref{lem:UX} we see that the square of the decay rate of
$\widetilde\Omega_{0x}^{+}$ exceeds that of $\widetilde\Omega_{00.x}$, giving the required result.

\emph{Part~(b)}.
Because of the rates of the FM-OLS estimator  established in
Proposition~\ref{prop:Rnull} and the rate of $\widehat\Omega_{00.x}$  in
Part~(a), under the null hypothesis we have for $k<1/3$
\begin{align*}
&W_I=
T^{2k}T^2\delta(T)^{-2} tr\left\{
\left(T^{-2}X'X\right)
\delta(T) \left(\widehat A^{+}-A^{0}\right)'
\left(T^{2k}\widehat\Omega_{00.x}\right)^{-1}
\delta(T)\left(\widehat A^{+}-A^{0}\right)
\right\}\\
&=
O_p\left(T^{2+2k}\min(T^{1+2k},T^{3/2})^{-2}\right)
=O_p\left(T^{-2k}\textbf{1}_{0<k\le1/4}+T^{2k-1}\textbf{1}_{1/4<k<1/3}\right)
=o_p(1),
\end{align*}
where $\textbf{1}_{A}$ is the indicator of $A$. 
\end{proof} 

\subsection{Proofs of the Theorems}
\begin{proof}[Proof of Theorem \ref{thm:RR}]
We have 
\begin{equation*}
\left[
\begin{array}{c}
B_{f.x} \\
0\\
B_{x}%
\end{array}%
\right]
=
\LL_{R}\left[
\begin{array}{c}
B_{0.x} \\
B_x%
\end{array}%
\right]
\equiv
BM(\LL_R \LL_{\Omega}\Omega\LL'_{\Omega}\LL'_R)
=
BM\left(\left[
\begin{array}{ccc}
\Omega_{ff.x}  & 0 &  0 \\
0 & 0 & 0 \\
0 & 0 &\Omega_{xx}%
\end{array}%
\right]\right),
\end{equation*}
where
\begin{equation*}
\LL_R=\left[\begin{array}{cc}
R' &  0 \\
R'_{\perp} &  0 \\
0 & I_{m_x}%
\end{array}%
\right].
\end{equation*}
The matrix $\left(R,R_{\perp}\right)$ rotates $u_{0.x,t}$ to
$(u'_{f.x,t},u'_{s.x,t})'$, where $u_{f.x,t}=R'u_{0.x,t}$ is $I(0)$ and
$u_{s.x,t}=R'_{\perp}u_{0.x,t}$ is $I(-1)$.
Therefore  $\LL_R \LL_{\Omega}$  keeps the nonstationary regressors $x_t$ and transforms the original cointegration relationship $y_{t}=A
x_t+u_{0t}$ to a system of two equations with orthogonal long run errors: (i)
an equation with $I(0)$ errors that has a nonsingular long run variance matrix $\Omega_{ff.x}$, and for this equation,
$R' y^{+}_{t}=R'A x_t+u_{f.x,t}$, Proposition~\ref{prop:Rfull} applies; and (ii) an equation with $I(-1)$
errors $R'_{\perp} y^{+}_{t}=R'_{\perp}A x_t+u_{s.x,t}$, for which Proposition~\ref{prop:Rnull} applies. 
\end{proof}

\begin{proof}[Proof of Theorem \ref{thm:Wald}]

Using coordinate rotation, we write the Wald statistic as a sum of several
components, corresponding to the nondegenerate and degenerate directions and
their cross products.
Recall the  partitioned matrix inversion formula
\begin{align*}
{\begin{bmatrix}
A_{11} & A_{12}\\
A_{21} & A_{22}\\
\end{bmatrix}}^{-1}
=
{\begin{bmatrix}
A_{11.2}^{-1} & -A_{11}^{-1}A_{12}A_{22.1}^{-1}\\
-A_{22.1}^{-1}A_{21}A_{11}^{-1} & A_{22.1}^{-1}\\
\end{bmatrix}},
\end{align*}
where the Shur complement is defined as
$A_{ii.j}=\left(A_{ii}-A_{ij}A_{jj}^{-1}A_{ji}\right)$. 
We apply the above formula to the variance matrix metric in the Wald test
statistic giving
\begin{align*}
\left(\left( R,  R_{\perp}\right)'
\widehat\Omega_{00.x}
\left( R,  R_{\perp}\right)
\right)^{-1}
&=
{\begin{bmatrix}
 R'\widehat\Omega_{00.x}R & R'\widehat\Omega_{00.x}R_{\perp}\\
R'_{\perp}\widehat\Omega_{00.x}R & R'_{\perp}\widehat\Omega_{00.x}R_{\perp}\\
\end{bmatrix}}^{-1}\\
&=
{\begin{bmatrix}
 \Omega_{ff.x}^{-1}+o_p(1) & O_p(1)\\
O_p(1) & -K^2\Omega_{e,e}^{-1}/w''(0)+o_p(K^{2})\\
\end{bmatrix}},
\end{align*}
where we take into account that
\begin{enumerate}[(i)]
\item $R'\widehat\Omega_{00.x}R\to_p\Omega_{ff.x}$, because $R$ isolates the nondegenerate
direction;
\item $R'_{\perp}\widehat\Omega_{00.x}R= O_p(K^{-2})$, which is obtained similar to
the proof of Lemma~1(b),
and switching from residuals $\widehat u_t$ to errors $u_t$ involves differences of order $O_p(K/T)$, as in Lemma~\ref{lem:0x}, which are of smaller order for $k<1/3$;
\item $K^2 R'_{\perp}\widehat\Omega_{00.x}R_{\perp}\to_p
-w''(0)\Omega_{e,e}$, which can be obtained similar
to the proofs of Lemma~1(c)  and Proposition~\ref{prop:RnullDenomDistr}(a).
\end{enumerate}
Then
\begin{align*}
&W_I=
tr\left\{
\left(T^{-2}X'X\right)
\left(\widehat A^{+}-A^{0}\right)'
\left(T R,  T R_{\perp}\right)
\left(
\left( R,  R_{\perp}\right)'
\widehat\Omega_{00.x}
\left( R,  R_{\perp}\right)
\right)^{-1}
\left(T R,  T R_{\perp}\right)'
\left(\widehat A^{+}-A^{0}\right)
\right\}\\
&=
tr\left\{
\left(T^{-2}X'X\right)
\left(\widehat A^{+}-A^{0}\right)' T R
\left(
 R' \widehat\Omega_{00.x} R
\right)^{-1}
T R' \left(\widehat A^{+}-A^{0}\right)
\right\}\\
&\quad{}+
T^{2+2k}\delta(T)^{-2}
tr\left\{
\left(T^{-2}X'X\right)
 \delta(T) \left(\widehat A^{+}-A^{0}\right)'R_{\perp}
\times O_p(1) \times
 \delta(T) R_{\perp}'
\left(\widehat A^{+}-A^{0}\right)
\right\}\\
&\quad{}+
T\delta(T)^{-1}
tr\left\{
\left(T^{-2}X'X\right)
\delta(T) \left(\widehat A^{+}-A^{0}\right)'R_{\perp}
\times O_p(1) \times
  T R'
\left(\widehat A^{+}-A^{0}\right)
\right\}\\
&\quad{}+
T\delta(T)^{-1}
tr\left\{
\left(T^{-2}X'X\right)
  T \left(\widehat A^{+}-A^{0}\right)'R
\times O_p(1) \times
 \delta(T) R_{\perp}'
\left(\widehat A^{+}-A^{0}\right)
\right\}\\
&= \chi^2_{r m_x}
+O_p\left(T^{2+2k}\delta(T)^{-2}\right)
+O_p\left(T\delta(T)^{-1}\right),
\end{align*}
using the fact that $T R'\left(\widehat A^{+}-A^{0}\right)=O_p(1)$
and $\delta(T) R_{\perp}'\left(\widehat A^{+}-A^{0}\right)=O_p(1)$ from 
Theorem~\ref{thm:RR}. For $k<1/3$, we have
$O_p\left(T^{2+2k}\delta(T)^{-2}\right)=o_p(1)$, and therefore $W_I\to_d \chi^2_{r
m_x}$.
\end{proof}

\end{document}